\documentclass[num-refs]{wiley-article}

\usepackage[utf8]{inputenc}
\usepackage[english]{babel}
\usepackage[normalem]{ulem}
\usepackage[autostyle]{csquotes}
\usepackage[backend=biber, sorting=nyt, style=authoryear, maxcitenames = 2, uniquelist=minyear, maxbibnames=99, giveninits=true, uniquename=init, doi=true, eprint=false, url=false, isbn=false, natbib=true, date=year]{biblatex}
\renewbibmacro{in:}{} 
\usepackage[breaklinks, hidelinks]{hyperref}
\usepackage{enumitem}
\usepackage{xpatch}
\usepackage{subcaption}
\usepackage{xcolor}
\usepackage{xspace}
\usepackage{multicol}
\usepackage[section]{placeins} 
\usepackage{times, amsmath, amssymb}
\usepackage{multirow,xfrac}
\usepackage{gensymb}
\usepackage{balance}
\usepackage[skip=.4\baselineskip plus0pt,indent=15pt]{parskip}
\usepackage{booktabs}

\renewcommand{\v}{\ensuremath{\mathbf{v}}\xspace}
\newcommand{\x}{\ensuremath{\mathbf{x}}\xspace}

\newcommand{\q}{\ensuremath{\mathbf{q}}\xspace}
\newcommand{\g}{\ensuremath{\mathbf{g}}\xspace}

\newcommand{\M}{\ensuremath{\mathbf{M}}\xspace}
\newcommand{\W}{\ensuremath{\mathbf{W}}\xspace}
\renewcommand{\P}{\ensuremath{\mathbf{P}}\xspace}
\newcommand{\Q}{\ensuremath{\mathbf{Q}}\xspace}
\newcommand{\V}{\ensuremath{\mathbf{V}}\xspace}

\newcommand{\J}{\ensuremath{\mathbf{J_g}}\xspace}
\newcommand{\D}{\ensuremath{\mathbf{D}}\xspace}

\newcommand{\R}{\ensuremath{\mathbf{R}}\xspace}

\renewcommand{\O}{\ensuremath{\mathcal{O}}\xspace}
\newcommand{\rs}{\ensuremath{R^2}\xspace}
\newcommand{\ros}{R\"ossler system\xspace}
\newcommand{\lst}{Lorenz~63 system\xspace}
\newcommand{\ross}{R\"ossler\xspace}
\newcommand{\lstt}{Lorenz~63\xspace}
\newcommand{\ie}{\textit{i.e.}\xspace}
\newcommand{\eg}{\textit{e.g.}\xspace}
\newcommand{\s}{Section~}         
\newcommand{\eq}{Equation~}       
\newcommand{\eqs}{Equations~}       
\newcommand{\fig}{Figure~}        
\newcommand{\figs}{Figures~}        
\newcommand{\tab}{Table~}         
\renewcommand{\st}{Step~}           

\usepackage[T1]{fontenc}
\usepackage[utf8]{inputenc}

\usepackage{color}

\usepackage{siunitx}

\papertype{Original Article}

\title{Supervised machine learning to estimate instabilities in chaotic systems: estimation of local Lyapunov exponents}

\author[1,2]{Daniel Ayers}
\author[3]{Jack Lau}
\author[1, 4]{Javier Amezcua}
\author[1,5]{Alberto Carrassi}
\author[3,6]{Varun Ojha}

\affil[1]{Department of Meteorology, University of Reading, Reading, UK}
\affil[2]{UK National Centre for Earth Observation, Reading, UK}
\affil[3]{Department of Computer Science, University of Reading, Reading, UK}
\affil[4]{Tecnologico de Monterrey, Campus Ciudad de Mexico, Mexico City, Mexico}
\affil[5]{Department of Physics and Astronomy ``Augusto Righi'', University of Bologna, Italy}
\affil[6]{School of Computing, Newcastle University, UK}

\corraddress{Daniel Ayers}
\corremail{d.ayers@pgr.reading.ac.uk}

\fundinginfo{UK Engineering and Physical Sciences Research Council,  Grant Number: EP/N509723/1; UK National Centre for Earth Observation Grant Number: NCEO02004; Schmidt Futures.
}

\runningauthor{Ayers et al.}

\begin{document}

\maketitle
\begin{abstract}
In chaotic dynamical systems such as the weather, prediction errors grow faster
in some situations than in others. Real-time knowledge about the error growth
could enable strategies to adjust the modelling and forecasting infrastructure
on-the-fly to increase accuracy and/or reduce computation time. For example one
could change  the ensemble size, or the distribution and type of
target observations, etc. Local Lyapunov exponents are known indicators of the
rate at which very small prediction errors grow over a finite time interval.
However, their computation is very expensive: it requires maintaining and
evolving a tangent linear model, orthogonalisation algorithms and storing large
matrices.

In this feasibility study, we investigate the accuracy of supervised machine
learning in estimating the current local Lyapunov exponents, from input of
current and recent time steps of the system trajectory, as an alternative to the
classical method. Thus machine learning is not used here to emulate a physical
model or some of its components, but ``non intrusively'' as a complementary
tool. We test four popular supervised learning algorithms: regression trees,
multilayer perceptrons, convolutional neural networks and long short-term memory
networks. Experiments are conducted on two low-dimensional chaotic systems of
ordinary differential equations, the \ross and the \lstt models. We find that on
average the machine learning algorithms predict the stable local Lyapunov
exponent accurately, the unstable exponent reasonably accurately, and the
neutral exponent only somewhat accurately. We show that greater prediction
accuracy is associated with local homogeneity of the local Lyapunov exponents on
the system attractor. Importantly, the situations in which (forecast) errors
grow fastest are not necessarily the same as those where it is more difficult to
predict local Lyapunov exponents with machine learning.


\end{abstract}

\section{Introduction}

Weather and climate are well known exemplars of chaotic dynamical systems. These
systems exhibit extreme sensitivity to initial conditions, meaning that initial
condition errors are subject to (on average) exponential growth until they reach
saturation~\citep{Lorenz1963, Kalnay2002}. The rate and the
characteristic of such growth, however, is highly state
dependent~\citep{Lighthill1986,Vannitsem2017}. As a consequence, although
chaotic systems have a finite predictability horizon (about $2$ weeks for the
atmosphere, see \eg~\cite{Holton2013}), the best estimate of prediction error
growth fluctuates in size along with the system's evolution, as the system goes
through periods of lower or higher predictability. For example, the short-term
predictability of the atmosphere depends on the weather regime present at a
given time \parencite[see, \eg][]{Palmer1996}.
Understanding the nature of error growth is essential to characterising a
system, and to enable better prediction.  The present work is
motivated by the idea that if the degree of
predictability of the system is known in real-time, it may be possible and
beneficial to take adaptive measures. For instance,  we speculate
that a local decrease of
predictability  might be counteracted by increasing the ensemble size in the context of ensemble-based
data assimilation or probabilistic forecasting, or the
distribution and type of target observations. Conversely, in areas of high
predictability, one  might save computational resources (and thus energy
consumption) via the opposite actions.  Understanding the impact of
such actions would require experimentation. In
this study we investigate the potential of machine learning (ML)
methods~\citep{Bishop1995,Hastie2009} to provide a real-time estimation of the system's local predictability. 

The mathematical theory of dynamical systems has long been the backbone to
understand and quantify predictability in deterministic chaotic systems. This is
commonly done by studying the instability properties of the solution, \ie by
analysing the linearised dynamics of small perturbations: the tangent space
evolution of these ``small'' perturbations are taken as proxies of the dynamics
of unknown initial condition errors \citep{ott_2002}. In this context, Lyapunov
exponents (LEs) are well-established quantities that measure the asymptotic
rates of error growth for a set of infinitesimally small errors that capture all
directions of phase space \citep{Pikovsky2016}. In practice, they measure the
average growth of small finite errors over long periods of time. The spectrum of
LEs is characteristic of each given dynamical system. Lyapunov exponents, and
their corresponding Lyapunov vectors (LVs), have been exploited in geosciences
for more efficient uncertainty quantification in data
assimilation~\parencite[\eg,][]{Palatella2013lyapunov,  Quinn2020, Albarakati2021,
Carrassi2022}, or for initialising probabilistic predictions
\parencite[\eg,][]{Toth1997, buizza2019introduction, Vannitsem2020}. The LE spectrum
can also be used to calculate other characteristic properties of a system, such
as the Kolmogorov-Sinai entropy, measuring the rate of information loss
\citep{Sinai2009}, or the Kaplan-Yorke attractor dimension \citep{Kaplan1979}.

We note that LEs are associated with directions known as
covariant Lyapunov vectors (CLVs). CLVs also provide useful information and can
be calculated numerically (see \cite{Ginelli2007, Wolfe2007, Froyland2013}).
 However, in
this work, we focus on the  exponents only.

The LEs are calculated as an average of finite-time Lyapunov exponents which are
here referred to as {\it local Lyapunov exponents}, LLEs
\citep{Benettin1980theory, Benettin1980numerical_application, Kuptsov2012}.
Whereas LEs provide ``global'' information about the average growth of small
perturbations in the system, the LLEs describe ``local'' growth rates along a
finite-time section of the trajectory. Notably, the LLEs show the heterogeneity
of the instabilities in phase space: the fluctuation of the local dynamical
stability around the asymptotic value as the system state varies
\citep{Sandri1996, Pikovsky2016}. This makes the LLEs ideal quantities to
measure the local degree of predictability, yet a bottleneck for their real-time
use in operational scenarios is the huge computational cost. Computational cost
grows quickly with the system's dimension making it prohibitive even for
moderate size models, let alone for models as large as those currently used in
numerical weather predictions ($\mathcal{O}(10^9)$ dimensions). Using the
standard method \citep{Benettin1980theory, Benettin1980numerical_application,
Kuptsov2012, Pikovsky2016}, calculating the LLEs and LEs involves computing a
long trajectory of the system (including a spin-up needed to ensure the solution
has reached the model attractor), propagating perturbations (as many as the
number of desired LLEs) with the tangent linear model (\ie, the resolvent of the
model Jacobian), and then repeatedly performing a process of orthogonalisation
(\eg using a QR decomposition algorithm).

Despite the computational bottleneck, Lyapunov methods (\ie computing the
local and global LE spectrum and aforementioned associated properties, or
the Lyapunov vectors) have been used for
dynamical
analysis of geophysical models of intermediate order ($\O(10^3)$ to $\O(10^5)$
variables), for example see \textcite{Vannitsem2016, Vannitsem2017,
DeCruz2018}. Additionally, Lyapunov methods have been applied to weather reanalysis
data to analyse the dynamics of the North Atlantic Oscillation
\citep{Quinn2021} and of persistent states of atmospheric pressure over the
European and western Asian continents \citep{Quinn2022}. In these works, the
bottleneck was overcome by reducing the data dimension (using empirical
orthogonal functions) and constructing
a reduced model. 
Whilst these works demonstrate the utility of Lyapunov methods, they do not
provide a means of calculating LLEs that is cheap enough to be carried out regularly
during a forecasting cycle.
    
Avoiding the need for such a costly computation whilst attaining an estimate of
the LEs or LLEs has thus great relevance. 
In their recent work, \textcite{Chen_et_al21} show how the outcomes of properly tuned data assimilation experiments can reveal the first LE as well as the Kolmogorov-Sinai entropy of the underlying dynamical model. The present work also seeks to avoid the cost of classical calculation methods, albeit only when time is critical. We investigate the feasibility of using certain ML methods \citep{Bishop1995, Hastie2009} to estimate the LLEs based only on information from the system's solution. Our focus is on supervised learning, which uses a data set of input-output pairs. The targets, \ie the desired outputs, are LLEs calculated using the classical method of evolving perturbations via the tangent linear model and orthogonalising. In this way, the cost of such methods is paid during the training phase of the ML method, and is avoided when making predictions. 

In the area of weather and climate forecasting, supervised learning has been used for various purposes \parencite[see \eg,][and references therein]{ Reichstein2019, Rasp2020bench, Chantry2021, Duben2021techmemo}. These include: (i) to emulate the full dynamics of a system \citep{Pathak2017, Pathak2018, Fablet2018, Nguyen2019, brajard20, Patel2021, Schultz2021, Sonnewald2021}, (ii) to improve a physics-based model with data-driven correction or parameterisation \citep{OGorman2018, Rasp2018, Bolton2019, Rasp2020, Bonavita2020, Nguyen2020, Gottwald2021, Brajard21}. Both approaches imply an intervention on the original model: the first approach yields surrogate data-driven models of the full original system, while the second approach builds hybrid models with data-driven and physics-based components. In either case the spectrum of the LEs of these new models can be computed using the standard approach \citep{Benettin1980theory, Benettin1980numerical_application, Kuptsov2012}, and can be compared to that of the original model as a way to quantify the goodness of the ML-reconstructed dynamics \citep{Pathak2017,brajard20}. 

In contrast to these two families of methods, this study aims towards improving
prediction skills by equipping the model with an external tool to quantify in
real-time the local degree of predictability, and thus guide ``non intrusive''
adaptations whereby the model equations are left unaltered. More specifically,
the goal is to use ML to predict the current LLE spectrum given input of the system state at the current and (possibly) most recent
time steps. We envisage that the trained ML algorithm could then be interrogated
for information about the local dynamical instability whilst performing the
numerical model forward integration.  We speculate that such information could drive a decision
process for adaptive modelling, for example adjusting the ensemble size when doing
ensemble-based data assimilation or probabilistic predictions,
 changing the distribution and type of target observations,
or adapting the numerical integration
scheme. Such adaptations could  mitigate error, improve
uncertainty quantification, or reduction computational cost.

In this feasibility study, we test the accuracy of some popular supervised ML algorithms in this task in two prototypical low-dimensional chaotic dynamical systems. This study is concerned solely with the predictive capability of ML methods: the task of optimising the computational cost of making predictions is left for future work. We anticipate that the latter task will be largely dependent on the specific use-case and computing hardware. The ML algorithms we test are regression trees (RTs) \citep{breiman1984}, multilayer perceptrons (MLPs) \parencite[\eg see][Chapter~6]{Goodfellow-et-al-2016}, convolutional neural networks (CNNs) \citep{LeCun1990}, and long short-term memory networks (LSTMs) \citep{Hochreiter1997,Graves2012}. These algorithms encompass three approaches to exploiting the temporal structure of the input. We evaluate both their point-wise accuracy and their statistical performance, measured in this case by the closeness of the distribution of predictions to the distribution of the target values. 
We find that on average the machine learning algorithms predict the stable local Lyapunov exponent accurately, the unstable exponent reasonably accurately, and the neutral exponent only somewhat accurately.
Each exponent is predicted more accurately in the \lst than in the \ros.
We show that greater prediction accuracy is associated with local homogeneity of the local Lyapunov exponents on the system attractor.
Importantly, the situations in which (forecast) errors grow fastest are not necessarily the same as those where it is more difficult to predict local Lyapunov exponents with machine learning.

The rest of this paper is organised as follows. In \s\ref{section LE theory}, we briefly review the theory of LEs and detail the method used to compute them. In \s\ref{section ML to estimate LLEs} we pose and conceptualise the ML problem we intend to solve, motivate the choice of the algorithms, and detail the input and target data and the evaluation metrics. In \s\ref{section system characteristics} we present the two systems under consideration: the R\"ossler and the Lorenz 63 models and discuss the characteristics of their Lyapunov spectra.  In \s\ref{section results} we present the results and \s\ref{section discussion} concludes with a discussion.

\section{Lyapunov exponents}\label{section LE theory}

\subsection{Overview of the theory}
We review briefly the theory of LEs, with the aim of providing an
intuitive explanation of what they are. 
The section follows \textcite{Legras1996, Benettin1980theory, Benettin1980numerical_application, Kuptsov2012, Pikovsky2016} and \cite[][Chapter~9, \s~3]{Strogatz2018}, to
which we refer the reader for a more rigorous and comprehensive treatment. 

Consider a deterministic autonomous dynamical system 
\begin{align}\label{eqn evolution}
    \dot{\x}=\g(\x),
\end{align}
where $\x\in\mathbb{R}^n$ is the state of the system, $\mathbf{g}\colon \mathbb{R}^n\to\mathbb{R}^n$ is the evolution function, and $\dot{\x}$ denotes
the derivative of $\x$ with respect to time. 
A trajectory of the
dynamical system starting at initial condition $\x(0)$ is  a set $\{\x(t)\colon
t\in A\}$, where $A$ is a connected subset of $\mathbb{R}_{\geq 0}$ containing
$0$. Consider the difference $\v(t)$ between a trajectory
started from the ``true'' initial condition
$\x(0)$, and a trajectory started from the perturbed initial condition  $\x(0)
+\v(0)$, where $\v(0)$ is infinitesimally small. 
The idea behind LEs is to find $\lambda(t)$, where 
\begin{subequations}
\begin{align}\label{eqn define lambda t}
    e^{t\lambda(t)} &= \frac{\|\v(t)\| }{\|\v(0)\| }\\
   \label{eqn define lambda t rearranged} \Leftrightarrow \lambda(t) &= t^{-1}\ln\left(\frac{\|\v(t)\|}{\|\v(0)\|}\right).
\end{align}
\end{subequations}
In other words, $\lambda(t)$ is the exponential growth rate of the initial error. 
In this setting, $\lambda(t)$ is specific to the initial condition $\x(0)$
and perturbation $\v(0)$, and $\lambda(t)$ varies with time. 

Lyapunov exponents generalise this notion to describe a) average
exponential growth rates of the system, regardless of the initial condition, and
b) the exponential growth rates for infinitessimal perturbations in
all directions.
To account for all directions, we consider perturbations
contained in an $n$-sphere of infinitesimal radius. As time progresses,
perturbations within the sphere are mapped into an ellipsoid. The LEs are the time-average of the exponential growth rate of the ratios
between the axes of the sphere and of the ellipsoid \citep{Legras1996}. 

Fix an initial condition $\x(0)$ and let $\v(0)$ be an infinitesimally small
perturbation, as above.
Then the dynamics of the perturbation are given by
\begin{align}
    \dot{\v}=\J \v , 
    \label{eqn perturbation}
\end{align}
where $\J$ is the Jacobian of $\g$ evaluated at $\x(t)$, \ie the linearisation of the evolution function at $\x(t)$. The solutions to \eq\ref{eqn perturbation} can be found using a
fundamental matrix, \ie any matrix-valued function  $\M(t)$ satisfying
\begin{align}
    \dot{\mathbf{M}}=\J \M\label{eqn fundamental matrix} ,
\end{align} 
and such that $\M(t)$ is non-singular for all $t$.  Focusing on a single perturbation, it follows from \eq\ref{eqn define lambda t} that
\begin{align}\label{eqn single perturbation LE}
    e^{\lambda(t)} = \left(\frac{\|\v(t)\|}{\|\v(0)\|} \right)^{\frac{1}{t}}
    = \left(\frac{\sqrt{\v(0)^{\rm{T}}\M(t)^{\rm{T}}\M(t)\v(0)}}{\sqrt{\v(0)^{\rm{T}}\v(0)}}\right)^{\frac{1}{t}}.
\end{align} 
We are interested in the value of $\lambda(t)$ as $t \to \infty$. Rearranging
\eq\ref{eqn single perturbation LE} and taking the limit, we have:
 
\begin{align}
\lambda=\lim_{t\to\infty} \ln\left[
\left(\v(0)^{\rm{T}}\M(t)^{\rm{T}}\M(t)\v(0)\right)^{\frac{1}{2t}}\right]. \label{eqn show naive limit}
\end{align}

 For almost all choices of $v(0)$, the $\lambda$ given by
\eq\ref{eqn show naive limit} is the largest LE.

We consider now the full spectra of LEs $\lambda_i(t),\ i=1,\dots, n$ that arise when one considers a sphere of
perturbations.  
The growth of a sphere of perturbations depends only on 
$\M(t)^{\rm{T}}\M(t)$. 
Thus we consider the limit
$\W(\x(0))$ defined by
\begin{align}\label{eqn define limit W}
   \W(\x(0))= \lim_{t\to\infty}\left[\M(t)^{\rm{T}}\M(t)\right]^{\frac{1}{2t}}.
\end{align}
By the multiplicative ergodic theorem \citep{Oseledets1968,
Ruelle1979} the limit exists, depends on the
initial condition $\x(0)$, and importantly, the eigendecomposition of
$\M(t)^{\rm{T}}\M(t)$ in the limit exists, which gives
\begin{align}
    \W(\x(0)) = \P(\x(0))\D \P^{\rm{T}}\!(\x(0)),
\end{align} where the eigenvector matrix $\P(\x(0))$ is orthonormal. The matrix of eigenvalues $\D$ is unique and depends neither on $\x(0)$ nor on the norm of the vector space containing the perturbations \citep{Kuptsov2012}. The LEs, $\lambda_1\geq\lambda_2\geq\dots\geq\lambda_n$, are the natural logarithm of the diagonal elements of $\D$. 

We finish with some remarks on the significance of the LEs. 
A chaotic system is a system with at least one positive LE. 
The LEs, above defined in terms of the growth of axes of a sphere of perturbations, are linked to the growth of the volume of the $n$-parallelepiped defined by the principle axes of the resulting ellipsoid \parencite[see][]{Wolf1985}. 
Also, the sum of the LEs is equal to the average divergence of the flow \parencite[see][\s2.5.4]{Pikovsky2016}. Thus, in dissipative systems, the sum of the LEs is negative. Finally, continuous chaotic systems have at least one LE equal to zero. This is due to there being zero growth of an infinitesimal perturbation in the direction of the flow. 

\subsection{Computation of local and global Lyapunov exponents}\label{subsec computation method}
The theory does not translate directly to a method for calculation of the LEs,
since in order to approach the limit in \eq\ref{eqn define limit W} one must
integrate \eq\ref{eqn fundamental matrix} to find $\M(t)$ for very large $t$.
This both accumulates numerical errors and results in a range eigenvalues of
$\M^{\rm{T}}\!(t)\M(t)$ that is too large for accurate numerical calculation
\citep{Pikovsky2016}.  Instead, we measure the growth of
perturbations over (finitely) many small time intervals, and compute the
average. Specifically, for each time interval we calculate the LLEs:
the natural logarithm of the growth ratios, divided by the length of the
time interval, as shown in \eq\ref{eqn define lambda t rearranged}. If the
system is ergodic, the arithmetic mean of the LLEs converges to the LEs as the
number of time intervals increases. Crucially, the perturbation vectors are orthogonalised and resized between each time interval. Orthogonalising the propagated perturbations is necessary to keep the perturbations distinct, since perturbations will tend to be attracted towards the direction of largest growth. The resizing is necessary to prevent perturbations becoming too small or large to be represented by floating-point numbers. We now present the algorithm used to calculate LEs and LLEs in this work, which is based on methods presented in \textcite{Benettin1980theory, Benettin1980numerical_application, Kuptsov2012}.

\begin{enumerate}
\item Calculate and store a long trajectory $\{\x(t)\colon t\in [0, T_{\rm{end}}]\}$. Discard an initial transient period to ensure the trajectory is in the attractor. One can alternatively calculate the trajectory at the same time as integrating \eq\ref{eqn fundamental matrix} in \st\ref{step iteration}\ref{step propagate perts} below, which avoids the need to store a long trajectory.

\item Initialise a matrix of perturbations $\Q_0 = [\mathbf{q}_0^1, \mathbf{q}_0^2, \dots, \mathbf{q}_0^n]$, such that the $\mathbf{q}_0^i\in{\mathbb R}^n$ are orthogonal and of unit length, that is, orthonormal.

\item\label{step iteration} Repeat the following iteration $m$
times, where $m$ is large enough to achieve convergence of the LEs. In iteration
$j$, starting with $j=1$, perturbations are propagated along the trajectory  
$\{\x(t)\colon t\in[(j-1)\tau, j\tau]\}$, where $\tau$ is typically small.
Each iteration results in $n$ LLEs: $\lambda_j^i,\ i=1,\dots, n$.
Henceforth we notate LEs with hatted lambdas to distinguish the asymptotic LE
$\hat{\lambda}_i$ from the LLE $\lambda_j^i$.
\begin{enumerate}
    \item \label{step propagate perts} Propagate the perturbations:
    $\V_j=\M(j\tau) \Q_{j-1}$, where $\Q_{j-1}$ is from iteration $j-1$, and
    $\M(j\tau)$ is computed by integrating \eq\ref{eqn fundamental matrix}.
    \item\label{step QR} Orthonormalise the propagated perturbations $\V_j$ to get $\Q_j$ using QR-decomposition \citep{Golub2013, Strang2016}: 
    \begin{align}
    \Q_j\R_j = \V_j.
    \end{align}
    \item  The diagonal elements $r_j^i$ of $\R_j$ are
    the desired ratios.
    The LLEs at time $j\tau$ are calculated as:
    \begin{align}\label{eqn compute LLE}
    \lambda_j^{i}=\tau^{-1}\ln(r_j^{\alpha(i)}),\ \ i = 1, \dots, n.
    \end{align}
     In \eq\ref{eqn compute LLE} the diagonal element $r_j^{\alpha(i)}$ is
    indexed by labelling function $\alpha(i)$, where $\alpha(i)$ is determined
    in \st\ref{step calc LEs}.
\end{enumerate}
\item \label{step calc LEs}Calculate the LEs: 
\begin{align}\hat\lambda_{i} = (m-k)^{-1}\sum_{j=k+1}^m \lambda_j^{i} \quad \quad\quad i=1,\dots, n
\end{align}
 where the LLEs from the first $k$ iterations are discarded. The bijective
 function $\alpha$ (in \eq\ref{eqn compute LLE}) takes inputs and values $1,\dots, n$, and is chosen such that
 the global LEs are numbered in descending order, \ie such that
 $\hat\lambda_{i}\geq\hat\lambda_{i+1}$. 
 We say $i^{\rm{th}}$ LLE to refer to any set
 of LLEs $\{\lambda_j^i\colon j=k,\dots,m\}$ that are associated to the $i^{\rm{th}}$ LE.
\end{enumerate}

In \st\ref{step calc LEs}, a transient period of length $k$ iterations is
required to let the initial perturbations $\Q_0$ converge to the dynamics of the
trajectory so that the leading perturbation $\mathbf{q}_j^1$ is oriented in the
direction of largest growth. As discussed above, the LLEs are defined in terms of ratios of the axes of the $n$-sphere and the
ellipsoid. In practice, it is
unlikely that the chosen initial perturbation $\q_0^1$ will be mapped by $\M(\tau)$
onto the leading axis of the ellipsoid, which will lead to a poor estimation of
the LLE. However, with sufficiently many iterations $k$, $\q_k^1$ will be
attracted to the direction of largest growth, leading to more accurate
estimates.

\paragraph{Computational cost}\label{sec computational cost}
The algorithm for computing the LLEs and LEs does not scale
well. The computational cost of the steps of the algorithm are as follows. The length of the required transient period, \ie the number of
"spin-up" iterations, depends on the system dynamics and, in the worst case,
grows proportionally to the system dimension $n$ \cite[p. 754]{Kuptsov2012}. For computing LEs, the total number of iterations $m$ depends on the complexity of the attractor and the precision required.  
Each iteration (\st\ref{step iteration}) requires the calculation of a trajectory of length $\tau$, which involves at least $\O(n)$ floating point operations (flops). The cost of integrating the matrix differential \eq\ref{eqn fundamental matrix} involves at least one evaluation of the Jacobian matrix, and at least one multiplication of the Jacobian by another matrix, per timestep. For a dense, non-trivial Jacobian, it is reasonable to assume that \st\ref{step propagate perts} involves $\O(n^2)$ flops in the simplest case, namely where $\tau=\Delta t$ and a minimal numerical integration scheme is used. In a less simple case, \st\ref{step propagate perts} will require at least $\O(n^{2.3})$ flops: the cost of multiplying two dense $n\times n$ matrices \citep{Alman2021}. \st\ref{step QR} is by far the most expensive step, since computing eigenvectors and eigenvalues via QR decomposition requires $25n^3$ flops \citep{Golub2013, Arbenz2016}. The overall theoretical time complexity of the LE algorithm is thus $25n^3 + \O(n^2)$ flops. In practice, $n \times n$ matrix multiplication can be much slower (at least $\O(n^3)$) due to memory access latency \citep{Albrecht2010}. 
Consequently, computing the full LLE spectrum for a modern weather prediction
system (where $n\approx 10^{9}$) is too expensive to be done during the forecast
cycle.

    Obviously, generating a subset of the LLE spectrum costs less. When using the tangent linear model, one
must compute consecutive leading LLEs: it is not possible to calculate LLE
$\lambda^i$ without also calculating $\lambda^1, \dots, \lambda^{i-1}$. The cost
of QR decomposition of an $n\times i$ matrix scales at $\O(ni^2)$
\citep{Boyd2018}. 
    The cost of multiplying an $n \times n$ matrix by an $n \times i$ matrix is
    $\O(n^2)$ when $i$ is sufficiently smaller than $n$ (in particular, at least
    when $\log_n(i)<0.31389$, see
 \cite{Huang1998, Christandl2020}). 
 In such cases,  
where $i\ll n$, the cost of
computing the LLEs scales as $\O(n^2)$ as it is dominated by matrix multiplication rather than
QR-decomposition. Use cases where it suffices to know a subset of the LLE
spectrum include assimilation in the unstable subspace (\eg see
\cite{Carrassi2022}) and computing the local Kaplan-Yorke dimension, which can
also be exploited for better data assimilation \citep{Quinn2020}.

\section{Using supervised machine learning to estimate \\ Lyapunov exponents}\label{section ML to estimate LLEs}

\subsection{Problem statement and evaluation metrics}\label{subsec problem
statement and eval metrics}
Supervised learning refers to ML algorithms that use data sets formed of input-target pairs, whereby the goal is to construct a statistical model that emulates the idealised function that maps from the input to the target. 
The input and target are multidimensional arrays of data, not necessarily of the
same dimensions.
A single input-target pair is known as an example; the size of a ML data set
refers to the number of examples it contains.

Supervised learning algorithms construct statistical models by optimising the
model's parameters using the data. 
In the problem of this study, the input is the system state at a set of
consecutive recent
time steps including the current time $t$. 
The target is the vector containing the full spectrum of $n$ LLEs calculated using the method described in
Section~\ref{subsec computation method} by integrating perturbations from time
$t-\tau$ to $t$. We choose to estimate the full LLE
spectrum, however we note that the ML approaches we use can be
easily adapted to the sub-problem of estimating a subset of the LLE spectrum
(henceforth, ``the sub-problem''),
such as the unstable and near-neutral LLEs. See also remarks in \s\ref{sec computational cost} and \s\ref{sec computation time}.
Generally, we have
\begin{align}\label{eqn define input target pair}
    (\text{input}, \text{target})=\left(\ (\x_{k-r}, \dots, \x_{k-1}, \x_{k})\,,\quad (\lambda_{k}^{1}, \dots, \lambda_{k}^{n})\ \right),
    \end{align}
where we recall that $\x_k\in{\mathbb R}^n$ denotes the system state at time
$k\Delta t$, and $\lambda_{k}^{i}$ is the $i^\text{th}$ LLE computed from the
interval $[k\Delta t - \tau, k\Delta t]$.

\paragraph{Point-wise accuracy} 
By point-wise accuracy we refer to the ability of a ML algorithm to predict a specific LLE at an arbitrary time $t$. For its evaluation we calculate a separate \rs score for each LLE in the spectrum, from a set of $d$ predictions and targets. The \rs score, also known as the
coefficient of determination, is given by
\begin{align}\label{eqn define R2 score}
    R^2\Big(\,\left\{\,(\ y_j,\hat{y}_j\ )\mid j=1,\dots, d\, \right\} \,\Big) = 1 - \frac{\sum\limits_{j=1}^d (y_j - \hat{y}_j)^2}{\sum\limits_{j=1}^d (y_j - \bar{y})^2} \in (-\infty, 1],
\end{align}
where, for each $j$, $y_j\in\mathbb{R}$ is the target output (\eg the $i$th LLE), $\hat{y}_j\in \mathbb{R}$ is the model's prediction, and  $\bar{y}$ is mean of the target outputs. In \eq\ref{eqn define R2 score}, the numerator is known as the sum of squares of residuals, and the denominator is known as the total sum of squares. An $R^2$ score of 1 is optimal, and an $R^2$ score of $0$ is as good as guessing the mean of the target values every time.

\paragraph{Similarity of prediction and target distributions}
In addition to the point-wise accuracy, we evaluate the statistical accuracy of the ML models with quantile-quantile (QQ) plots. QQ plots provide a simple non-parametric tool to compare the empirical probability
distributions generated by two samples \citep{Wilk1968}. In our case these are the predicted and the target values. To generate the plot, a set of quantiles (the $1000$-quantiles in our experiments) is computed for both samples. These quantiles are then plotted against each other in a scatter plot. If the two samples have the same empirical distributions, the scatter plot renders a 45 degree diagonal line (of course this is subject to sampling error, which diminishes as the sample size grows). Departures from this ideal result show differences in the location and scale parameters of the empirical distributions, as well as possible linear and non-linear relationships between the variables, see e.g. \textcite{NISThandbook} for a more detailed discussion. 
In our case, the QQ plots are useful to show which parts of the target distribution are well represented by the predictions.

\subsection{Supervised learning algorithms}\label{section SL algos}
We test four algorithms, summarised in \tab\ref{table sl
algos}, all well known in the ML community.  They are chosen to
represent commonly used, proven-successful supervised learning
algorithms. 
In this section we detail the algorithms, their structure and their
relative characteristics. The final details of the algorithms, including the
number of parameters, are determined by
hyperparameter tuning and described in \s\ref{section HPO}.
We note that superior performance in supervised learning tasks has been
achieved by conducting neural architecture search (NAS): an extensive (and costly)
optimisation of neural network (NN) 
architecture from a vast and highly flexible search space \citep{Zoph2018}. Here
we stop short of conducting such an NAS. Instead, we choose established architectures for four
distinct algorithms, and carry out hyperparameter optimisation for each, where
the hyperparameters include key architectural choices such as the number of
layers and the number of neurons in each layer.
We expect that the results from our selection give
a good indication of the possible performance of supervised learning in this
task.

\begin{table}[htbp!]
    \centering
    \resizebox{\textwidth}{!}{
\begin{tabular}{ll} 
    \toprule 
    \textbf{Algorithm}&\textbf{Architecture}\\
    \cmidrule(lr){1-1}\cmidrule(lr){2-2} 
    {Multilayer perceptron (MLP)} & One or more dense layers and one dense output layer  \\
    {Regression tree (RT)} & One tree per target LLE \\
    {Convolutional neural network (CNN)} & One 1D-convolution layer, max pooling
    layer, flatten layer, one or more dense layers\\
    {Long-short term memory network (LSTM)} & One or more LSTM layers, one dense output layer\\
    \bottomrule 
\end{tabular}}
\caption{ The four supervised learning algorithms used in this study. The hyperparameter values (\eg number of dense layers)
used in the experiment are described in \s\ref{section HPO}}
\label{table sl algos}
\end{table}

 As we discuss in the following paragraphs, the chosen algorithms
take different approaches to using the temporal structure of the input (when there are multiple time steps in the input). Here by temporal structure we mean the temporal sequence of elements
in each input vector, as opposed to the pairwise relation between the inputs and
outputs while segmenting the time-series for data set preparation. The relative
success of each
algorithm gives insight into the nature of the problem from the ML perspective.

 The first algorithm is the regression tree (RT)
\citep{breiman1984}. The RT is the only non-neural network algorithm we test; RTs function by evaluating a finite chain of comparisons on the input features, such as ``$x_k > 0$''.
The chain of comparisons form a tree graph; each leaf node corresponds to an output value.
Thus RTs have finitely many possible output values.
The key advantage of a RT is its low computational cost of
making predictions.
Other advantages include the implicit feature selection process and potentially
greater explainability of predictions compared with NNs.
In fact, the RT may make use of only a few features from the set available in the input vector~\citep{breiman1984}.
By setting two hyperparameters (the maximum number of leaf nodes and the maximum
depth, \ie number of consecutive comparisons before a leaf node), it is possible
to greatly constrain the size of the resulting tree.
For simplicity, we opt to train a separate
RT for each target LLE: \ie, the prediction of the target vector
$(\lambda_{k}^{1}, \dots, \lambda_{k}^{n})$ is made by $n$ RTs, where each RT
predicts a different $\lambda_k^i$.

 The second algorithm is the multilayer perceptron (MLP), the
most basic type of feedforward artificial NN,
\parencite[as described in, \eg,][Chapter~6]{Goodfellow-et-al-2016}. The MLP is
comprised of several hidden layers and an output layer, each of which is
comprised of many neurons, where each neuron receives as input the outputs from
all neurons in the previous layer (\ie, each layer is densely/fully connected). Each hidden layer has the same number of neurons;
this value is optimised as a hyperparameter (see \s\ref{section HPO}). When making a
prediction, the entire input vector is passed to every neuron in the first hidden
layer. These neurons compute their outputs according to their weights and
activation function, and their outputs are passed onto the neurons of the second
layer, etc., until the neurons in the final layer produce outputs which
are taken to be the final output of the algorithm. The MLP can theoretically
approximate any continuous function \parencite{Hornik1991}, however in practise
it has been found that in many problems accurate predictions are more easily
achieved with more sophisticated architectures.

With regards to using temporal structure, the RT and the MLP take the same
approach: to treat each 
element of the input vector (each feature) as an independent and identically distributed variable. 
Both algorithms have
no inherent preference with respect to the temporal structure of the data, since they are invariant
to the choice of order of the input features (a choice which is made in the preprocessing stage before training the
algorithm).

 The third algorithm is a convolutional neural
network (CNN) \citep{LeCun1990}.
In our experiments, the CNN is comprised of one 1D-convolution layer with a
kernel of size two and a stride of one, a
max-pooling layer (if the number of time steps in the input is greater than 2),
a flattening layer and finally a set of dense layers (the number of which is
optimised as described in \s\ref{section HPO}). 
The 1D-convolution layer is ``1D'' in that the layer convolves only over the time dimension. 
The kernel size of two means that if the input is comprised of system state
vectors at $r+1$ consecutive time steps, the convolution layer is
only sensitive to patterns that occur in any time window of length two within the
$r+1$ time steps.
The max-pooling layer has a
pool size and stride of two, which results in an invariance to translation
of patterns by a single time step.  Thus the CNN takes a different approach to
using the temporal structure of the input, by only being sensitive to patterns that appear in
small time windows within the input.  In other words, the convolution layer
predisposes the CNN to be aware which system states in the input are adjacent to
each other in time.  CNNs have shown to be very successful in a range of
applications, including in image-related tasks \citep{LeCun2015}. The CNN architecture in our experiments
is equivalent to the MLP with an additional 1D-convolution layer at the
beginning. We choose a 1D convolution (\ie a convolution along the time
dimension only), because in the ODE systems in our experiments, there are only three state
variables and there is not a spatial locality which would motivate a focus on
two of the variables at a time.

 The fourth algorithm is the long short-term memory network
(LSTM) \citep{Hochreiter1997,Graves2012}. 
An LSTM is a form of recurrent neural network (RNN): the NN processes data
in a sequence, and after each term in the sequence is processed, information is
stored in a hidden state. In an LSTM, the flow of information
in and out of the NN's hidden state is controlled by learned gates. 
LSTMs have been successful in various tasks where
there are long-term dependencies, \ie 
where the correct output at a later element in the sequence requires information
from elements further back in the sequence. 
Such tasks include speech recognition \citep{Graves2013} and machine translation \citep{wu2016googletranslate}. 
In machine translation, for instance, it is useful to retain information
about words early on in the sentence to best predict how to translate words at
the end of the sentence.  However, the sophistication of the LSTM architecture
comes at a computational cost which we noticed particularly during training. 
In the experiments of this study, the LSTMs are composed of  
The LSTM layers process the input one time step at a time, in chronological order.
The estimation of the LLEs is made after the last time step has been processed.
 Thus, the LSTM takes a third approach to using the temporal structure by using
information from past time steps when processing future time steps.

The NN algorithms (MLP, CNN and LSTM) were implemented using Tensorflow \citep{tf}. Additionally, all NN algorithms standardise the input and target data by subtracting
the mean and dividing by the standard deviation, where the mean and standard
deviation are calculated from the training set. 
Each component of the input and target is standardised independently. 
This is to make the data more amenable to learning.
The scaling pipeline was implemented using Scikit-learn \citep{scikit-learn}.
For the RT, we use the
implementation in the Scikit-learn Python module
\citep{scikit-learn}. 

\section{R\"ossler, Lorenz~63 and their local Lyapunov spectra}\label{section system characteristics}

In this section we present the two dynamical systems used in this work. We
discuss the characteristics of their attractors and their LLEs, as this forms
the data for the ML models and will be important in understanding their
performance. The \ross \citep{Rossler1976} and the \lstt \citep{Lorenz1963}
systems are both three-dimensional, continuous time, ODE dynamical systems,
given respectively by \eqs\ref{eqn Rossler}~and~\ref{eqn Lorenz}.  We use the parameters $(a,b,c) = (0.37, 0.2, 5.7)$ and $(\sigma, \rho, \beta) =
(10, 28, 8/3)$. These are commonly chosen values for which the systems exhibit chaotic behaviour \parencite[see \textit{e.g.}][]{ott_2002}. Under these settings, both systems are dissipative, and they possess strange attractors of fractal dimension.
\vspace{-10pt}
\noindent
\begin{center}
  \noindent
  \begin{minipage}[b]{.45\textwidth}
    \begin{align}\label{eqn Rossler}
      &\textbf{R\"ossler}\notag \\[-6pt]
      \openup -1\jot 
      \begin{split}
        &\dot x = -y -z\\
        &\dot y = x+ay \\
        &\dot z = b + z(x-c)
        \end{split}
    \end{align}
  \end{minipage}
  \quad
  \begin{minipage}[b]{.45\textwidth}
    \begin{align}\label{eqn Lorenz}
      &\textbf{\lstt}\notag\\[-6pt]
      \openup -1\jot 
      \begin{split}
           &\dot x = \sigma(y-x)\\
          &\dot y = x(\rho-z)-y\\
          &\dot z = xy-\beta z
      \end{split}
       \end{align}
  \end{minipage}
\end{center}

 The \lst is famous amongst weather and climate scientists: initially derived as
a truncated model of Rayleigh-B\'{e}rnard convection in a two-dimensional fluid flow, it is the archetypal chaotic system
and continues to be used in weather and climate science, for example in data
assimilation experiments \parencite{Carrassi2018}. For the chosen system
parameters, 
the attractor of the \lst is formed of two wings, each centred
around a non-stable fixed point. There is a third non-stable fixed
point at the origin \parencite{Sparrow1982}. 
The \ros was introduced as a simpler version of the \lst, having only one
non-linear term ($zx$) instead of the two in the \lst ($xz$ and $xy$). 
\eqs\ref{eqn Rossler} were derived as a simplification
of a system that combined two chemical reactions: a slow, two-variable
oscillator ($x$ and $y$) and a faster ``switching-type'' reaction ($z$) \parencite{Rossler1976}. For a historical review of the development of the \ros,
see \textcite{Letellier2010}. The resulting system attractor is composed of a
``disc'' in the $xy$ plane ($z$ close to $0$), and a ``loop'' in which trajectories
rise ($z$ increases rapidly) out of the disc, before folding over and back into
the disc. Consequently there is an imbalance: any infinite, non-periodic trajectory spends more time
in the disc than in the loop. The attractor has two fixed points: an unstable fixed point in
the
centre of the disc, around which system trajectories spiral outwards, and a
stable fixed point located outside of the attractor. We shall see that the difference in the
systems' dynamics provides useful comparison.

It is easy to compute the analytic Jacobian matrices from \eqs\ref{eqn Rossler}
and \ref{eqn Lorenz}. Furthermore, the small size of both systems allow us to
perform exhaustive experiments with long time series. 
The data used in our results are generated following \s\ref{subsec computation
method}. 
The
evolution equations, \eqs\ref{eqn Rossler} and \ref{eqn Lorenz}, as well as the
corresponding fundamental matrix equation, \eq\ref{eqn fundamental matrix}, are
integrated using a fourth-order Runge-Kutta scheme with timestep $\Delta t =
0.01$ for both systems. 
The LLEs are calculated over time windows of length
$\tau = 0.04$ ({\it i.e.} four timesteps); the LLEs at time $t$ are calculated
by integrating perturbations from time $t$ to $t+\tau$. 
We choose  $\tau = 0.04$, rather than a smaller value, so that the resulting set
of LLEs provide a better coverage of the attractor.
\tab\ref{table LEs} shows the resulting LEs.  Further details of
the data used in the results are given in \s\ref{sec experiment setup}.

\begin{table}[htbp!]
\centering
 \begin{tabular}{c c c c}
 \hline
  System & $\hat\lambda_1$ &  $\hat\lambda_2$ & $\hat\lambda_3$ \\
 \hline
  \ross & $0.19597\,\pm\,0.00011$ & $0.0000075\,\pm\,0.0001275$  &$-4.30097\,\pm\, 0.00068 $\\ 
  \lstt & $0.90495\,\pm\,0.000145 $&$ 0.001975\,\pm\,0.000055 $ &$-14.571345\,\pm\,0.000105 $\\[0.5ex]
 \hline
 \end{tabular}
 
 \caption{The LEs of the \ross and \lstt systems as calculated following
 \s\ref{section LE theory}. These values are computed from 718,800 LLE
 iterations, following a transient of $k=1,200$ iterations. As described in \s\ref{section LE theory}, the
 arithmetic mean of the LLEs converges to the LEs as one includes LLEs from more
 iterations, \ie, from a longer trajectory. The convergence is not monotonic:
 the series of arithmetic means fluctuates as the number of iterations increases. To
 give an indication of the precision of the numerical calculation, we therefore
 calculate the LE as the mid-point of the range of the series of arithmetic
 means acquired from the first $j$ iterations, where $j = 716,801, \dots, \allowbreak
 718,800$. The extent of the range above and below this value are also given.
 The proximity of the second LE to zero is a test for the accuracy of the
 numerical algorithm, since the middle LE is theoretically known to be zero in
 chaotic autonomous continuous-time systems \citep{Pikovsky2016}.}
 \label{table LEs}
\end{table}

\begin{figure}[htbp!]
\centering
\includegraphics[width=0.62\linewidth, clip]{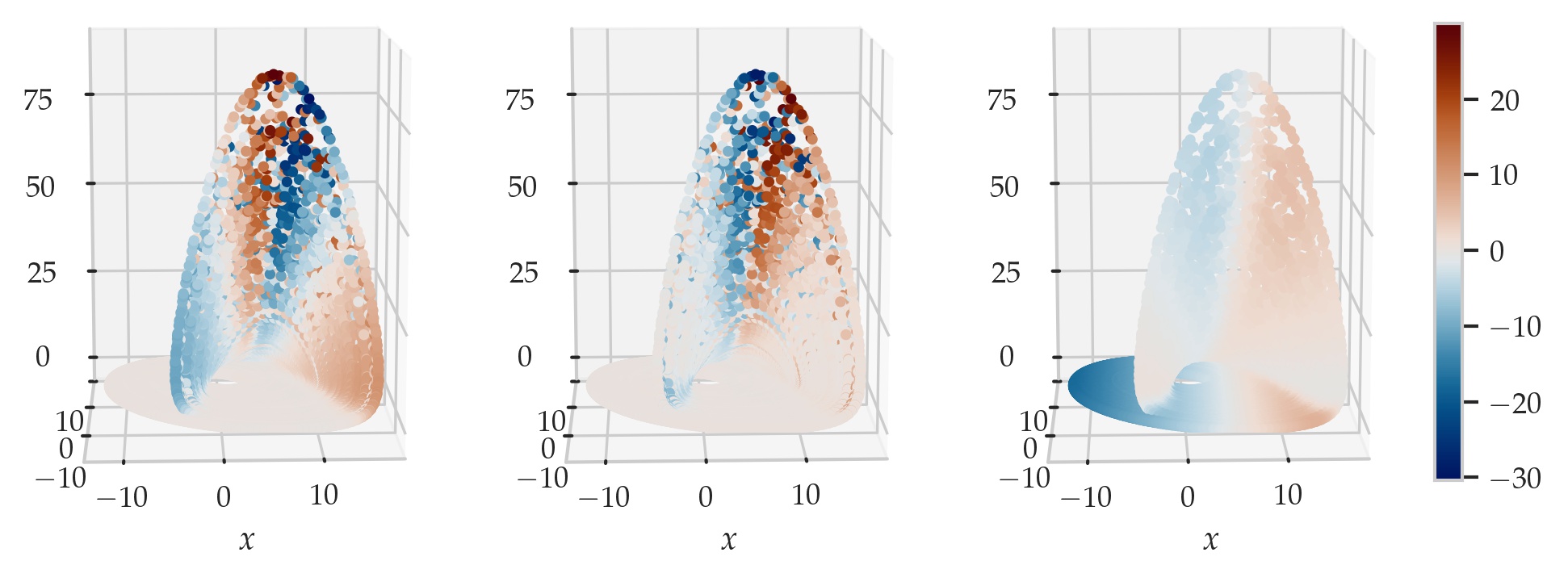}\\
\includegraphics[width=0.64\linewidth, clip]{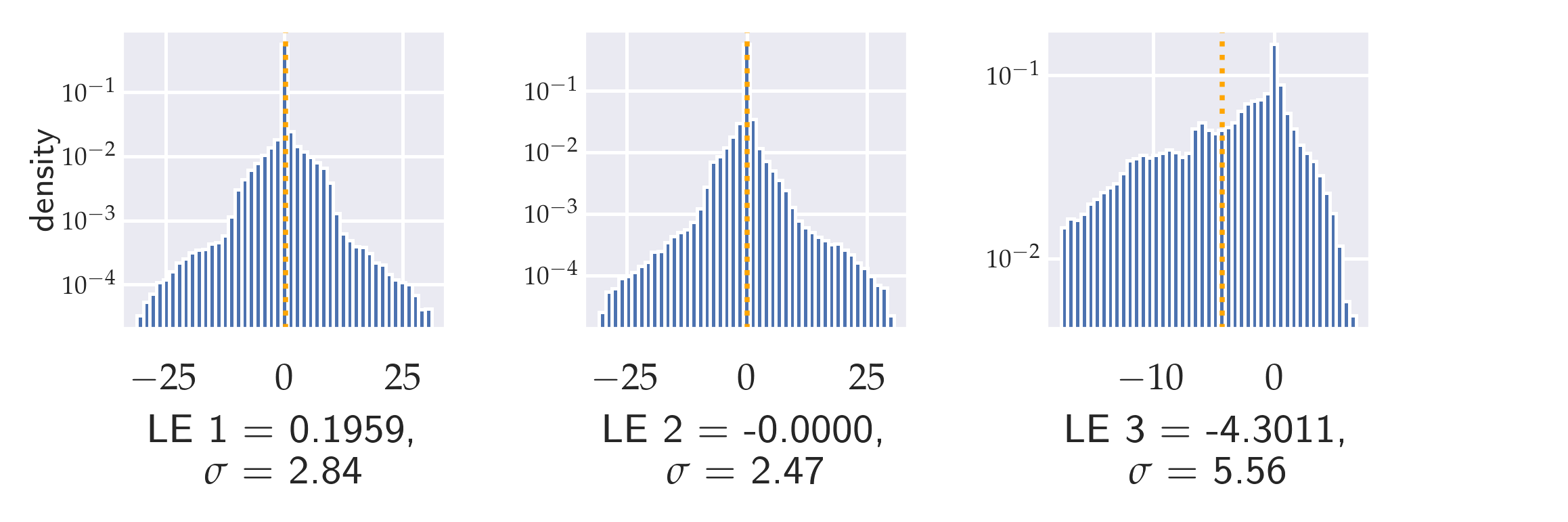}
\caption{A long-time trajectory of the \ros coloured by LLE values (top row) shows how LLE values tend to be arranged in the system's attractor. The bottom row shows the corresponding statistical distribution of the LLEs via histograms; the mean of the LLE values is shown by the dotted orange line. The top panels show the same collection of $25,000$ points. On the other hand, the histograms are generated from the full set of $718,800$ LLEs; note that the vertical axis is plotted in logarithmic scale. The mean (i.e. the corresponding LE) and standard deviation of the LLE values in each column are shown underneath.} 
\label{fig distributions of LLEs Rossler}
\end{figure}

In \figs\ref{fig distributions of LLEs Rossler} and \ref{fig distributions of
LLEs Lorenz}, the top row shows the values of the LLEs along the systems'
trajectories, specifically for the points $\x(\tau j)$ where $ 1,200\leq
j\leq 26,199$; the values of the LLEs are given in color. We show only $25,000$
data points to avoid saturating the figures: the local heterogeneity of the LLE values in phase space is evident. The bottom
row displays the distribution of LLE values via histograms, for the
points $\x(\tau j)$ where $ 1,200\leq j\leq 719,999$. The $i$th column
shows the LLEs associated with the $i$th LE. 

\begin{figure}[htbp!]
\centering
\includegraphics[width=0.62\linewidth, clip]{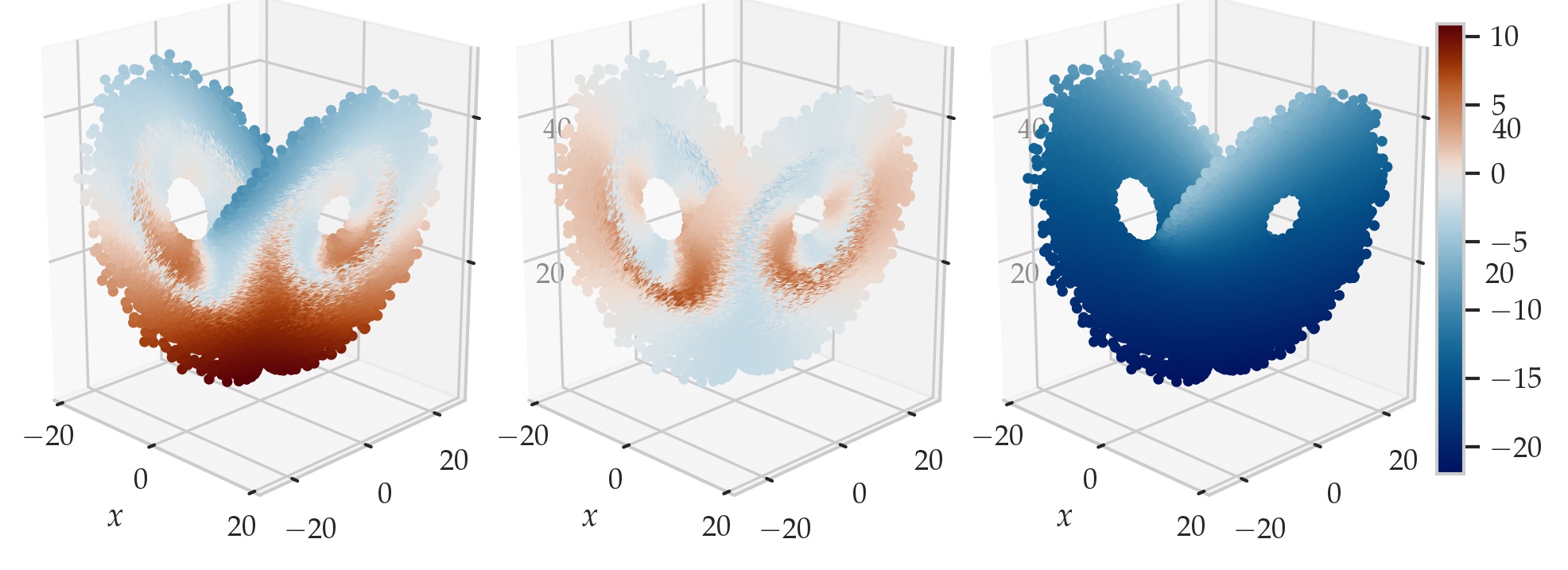}\\
\includegraphics[width=0.64\linewidth, clip]{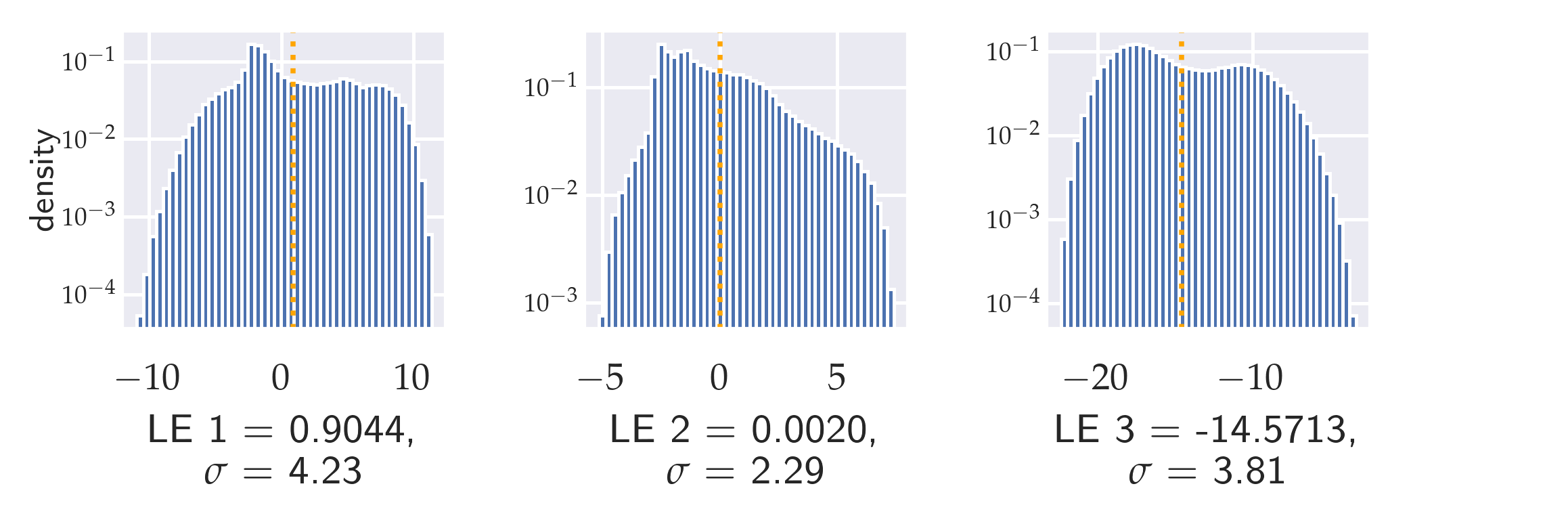}
\caption{As with \fig\ref{fig distributions of LLEs Rossler} but for \lstt.} 
\label{fig distributions of LLEs Lorenz}
\end{figure}

Let us first consider the statistical distribution of the LLEs. In the \ros (\fig~\ref{fig distributions of LLEs Rossler}) there is a marked difference between the first two LLE distributions and the third. The first two have a single, tall, thin mode with long tails and are roughly symmetric. The distribution of the third LLE has a lower-density mode with thicker tails and is negatively skewed. The range of values of the third distribution is also much smaller than that of the first two. 

The \lst LLE distributions differ from those of the \ros. All three LLE distributions (\fig\ref{fig distributions of LLEs Lorenz}) have thicker tails and are positively skewed. The first and third are weakly bimodal, whilst the second is unimodal.

We now introduce a property that will be important to understanding the
performance of the ML. In the context of the distribution of LLEs in the
attractor of a system,
we say that a region $U$ of the attractor $A$ is \textit{(locally)
heterogeneous} if 
the function $f\colon A \to \mathbb{R}$ mapping from
$A$ to the $i$th LLE value is non-smooth in $U$.
The best definition in more simple terms is that the LLE values in $U$ vary in a
non-smooth fashion across $U$.
For example, an alternating lattice (such as a chess board) would be locally
heterogeneous.
If $U$ is not heterogeneous, we say it is homogeneous.
In this work we use homogeneous and heterogeneous only in the sense of
\textit{locally across a small region of phase space $U$}, and not in the sense of across the
entire attractor (\ie globally), or along a trajectory (\ie across time), as is
the case in, \eg, \textcite{Vannitsem2017}, and \textcite{ lucarini2020new}.

For all three \ross LLEs, in the disc of the attractor that sits in the $xy$-plane,
the LLE values are homogeneous. 
In the loop that jumps out of the disc with positive $z$-values, there are bands
of similar values for all three LLEs. 
For the first two LLEs, the most extreme LLE values are in the loop, and there is
significant local heterogeneity within the bands. 
For example, there are some small regions of the attractor
where the first LLE is $10$ or above for the majority of points, yet for some other points it is as low as $-20$. 
This high degree of heterogeneity is reflected in 
the longer tails of the distributions of the first and second LLEs ({\it cf} \figs\ref{fig distributions of LLEs Rossler}~and~\ref{fig distributions of LLEs Lorenz}). In contrast,
the third LLE is locally homogeneous: the colour change in the graph is smooth. 
Note that the trajectory of the \ros spends
more time in the disc than in the loop. Consequently, data points in the loop
are considerably sparser.

In the \lst, the first two LLEs have distinct regions of local heterogeneity and
homogeneity. Unlike with the \ros, the regions of greater mixing are along the
boundaries between regions of greater local homogeneity. The most extreme values
of the first LLE are at the bottom of the attractor ($z$ close to 0) and on the
top edge of the wings: not in the regions of local heterogeneity. 
The third LLE, as with the \ros, is everywhere locally homogeneous.
As we will see, the characteristics of the LLEs we have discussed have implications for the performance of the ML methods trying to estimate them.

\section{Experiments and results}\label{section results}

\subsection{Experiment setup}\label{sec experiment setup}
Having described the ML algorithms (\s\ref{section SL algos}) and dynamical systems (\s\ref{section system characteristics}) used in this work, this section details the experiments and results. The experiments test a total of $16$ configurations, which consist of each combination of two dynamical systems, four ML algorithms, and two input types: 
\begin{equation}
\begin{aligned}
    \text{Input Type 1:} & &(\x_k)&\\ 
    \text{Input Type 2:} &  &(\x_{k-5},\,\x_{k-4},\,\x_{k-3},\,\x_{k-2},\,\x_{k-1},\,\x_{k})&\\
\end{aligned}
\end{equation}
The first input type consists of the
current time step only, whereas the second has the current time step and the five
preceding time steps.  In the envisaged operational application
of this approach, storing multiple time steps of the entire model state poses a
severe computational challenge. Since we envisage making LLE predictions every
time step, and we have assumed that the pattern of input time steps remains
fixed, the furthest-back input time step dictates the number of time steps
that must be stored. We choose $5$ time steps into the past as a balance between testing whether
previous time steps can enable more accurate predictions, and not requiring huge
amounts of steps to be stored. Ignoring the constraints of feasibility and our
approach, we expect that delays of more than $5$
time steps might
enable more accurate predictions. Given the choice of a maximum $5$ time steps
into the past, we include all $6$ time steps in the input and rely on the ML
algorithms to extract useful features thereof.
Note that if there is only one time step in
the input, the CNN is equivalent to an MLP where the number of units in the first hidden layer of the MLP is twice the number of filters in the CNN.

 The main results were attained with data sets of $10^5$ examples. The data sets were created
using the method
described in \s\ref{subsec computation method}, and the parameters given in
\s\ref{section system characteristics}.
For our chosen value of $\tau=0.04$, the $10^5$ examples are generated from a
trajectory of $4\times 10^5$ time steps, equating to $784$ and $3604$ Lyapunov
times for \ross and \lstt, respectively. We note that another study which
estimated the global LEs of the \lst by emulating the dynamics with reservoir computers
used a far smaller data set of $91$ Lyapunov times \parencite{Pathak2017}. By
inspecting plots of the trajectory (not shown here) we anticipated that $10^5$
examples provide sufficient coverage
of the system attractors. The Kullback-Leibler divergences of input and target
variables (not included here) show that $10^5$ examples provides a good
representation of the variables' statistical distributions (when compared to a
data set of $ 10^6$).  To assess the impact
of the data set size, we repeated experiments with data sets of $5\times 10^4$
and $10^6$ examples. We found that the much larger data set resulted in
only slightly more accurate predictions, with the exception of the LSTM. Thus
the initial choice of $10^5$ does not limit performance substantially. These results are discussed further in
\s\ref{Section results subsection}.

From each system's data set we generate $30$ data set instances, where each instance is a
unique random shuffle of the original.
Each data set instance is partitioned into training, validation and test sets with a
ratio of $0.6: 0.2: 0.2$.  The resulting data setup is summarised in
\tab\ref{table data setup}.
The training, validation and test sets of different data set instances are therefore
distinct.
Given that the \ros and \lst are both ergodic, and the size of the training, validation
and test data sets are large, the coverage of the attractors by examples in each
data set instance is similar to that which would be achieved if we had
instead generated $30$ sets of new data.
Each configuration is tested on the $30$ data set instances, \ie each
configuration is tested in $30$ trials.
This provides an estimate of the variability of performance of each algorithm.

\begin{table}[htpb!]
    \centering
    \resizebox{0.38\textwidth}{!}{
    \begin{tabular}{ll} 
        Number of data set instances&$30$\\
Number of examples:&\\
\,\,\, in each data set instance&$100,000$\\
\,\,\, in the training partition&$60,000$\\
\,\,\, in the validation partition&$20,000$\\
\,\,\, in the test partition&$20,000$
\end{tabular}}
    \caption{A summary of key values in the data setup  for the $10^5$ data sets. Note that each data set instance contains the same examples in a unique, shuffled order.}
    \label{table data setup}
\end{table}

The NNs (MLP, CNN, LSTM) are implemented with two methods for preventing overfitting:
activity regularisation on each layer, and early stopping. Regularisation
penalises large weight values; early stopping selects the model weights that
score optimally on the validation data set, rather than on the training data set.

\subsection{Hyperparameter optimisation}\label{section HPO}
In ML, {\it hyperparameter} refers to any parameter that has to do
with the form of, or method of optimising, the statistical model, as opposed to
the trainable parameters of the model itself (often referred to as weights).
Typically, hyperparameters are fixed before the model is fit to the data, \ie before the
weights are optimised.
For example,
one hyperparameter for a NN is the learning rate: the amount by
which model weights are changed at each step in the optimisation. Selecting the right hyperparameters
is essential for effective use of ML algorithms
\citep{Goodfellow-et-al-2016}. 

We use a Bayesian optimisation method, implemented by Scikit-Optimize \citep{scikit_optimize}, to optimise the hyperparameters for each
ML algorithm used in this study. Conceptually, a Bayesian optimisation method is an informed hyperparameter search that generates a probabilistic model (\eg, using a surrogate Gaussian process regression) of the true ML model (\eg, CNN) to select a set of hyperparameters that maximises the true ML model's performance \citep{Snoek2012}. 

We perform separate hyperparameter optimisation for each configuration. 
Hyperparameter
optimisation was
carried out on Google's tensor processing units (TPUs) using Google
Colaboratory. The search domain for each hyperparameter was
chosen based on users' knowledge of the algorithms, the nature of the problem, and common practice in the
ML community~\citep{Hastie2009,Goodfellow-et-al-2016}.
The search domains are shown in Table~\ref{table HPT search space}. The chosen search space permits the NNs to be reasonably large  (up to $200$
neurons per layer and up to $10$ layers for MLP and CNN, and up to 100 LSTM
units per layer and up to 3 LSTM layers for LSTM) given the low-dimensionality of the task: mapping from three ($3$
variables $\times\ 1$ time steps), or eighteen ($3\times 6$) features to three
outputs. 
On the other hand, we opted to restrict the maximum number of
layers for the LSTM due to the greater complexity of the algorithm. In fact, the first
attempt to perform hyperparameter optimisation with a $6$-layer LSTM exceeded the
$24$ hours runtime limit for Google Colaboratory. In contrast to the NNs, we forced the RTs to remain computationally light by limiting
\textit{maximum leaf nodes} (\ie the maximum number of possible output values) to $100$.
The
entire hyperparameter optimisation process had a combined runtime of $108$ hours.
Although the hyperparameter optimisation was computationally expensive it was affordable in the low dimensional problems at hand. It
 greatly increases the chance
that we attain maximal performance from each ML algorithm, thus providing
useful insights into the problem from a ML perspective.

\begin{table}[htb!]
    \centering
    \resizebox{0.8\textwidth}{!}{%
\begin{tabular}{lll} 
\toprule 
Algo&Hyperparameter&Search domain\\
\cmidrule(lr){1-1}\cmidrule(lr){2-2}\cmidrule(lr){3-3} 
\textbf{RT} & maximum depth & $[1,\ 100]$\\
& maximum leaf nodes&$[5,\ 100]$\\
& maximum features& $\{$None, $\log_2$, square root$\}$\\
&splitter& $\{$best, random$\}$\\
& min. cost-complexity pruning parameter&$[1 \times 10^{-6},\ 100]$ \\
& minimum examples per leaf & $[1,\ 20]$\\
&minimum weight fraction per leaf &$\{0, 0.5\}$\\
\cmidrule(lr){2-2}\cmidrule(lr){3-3}
\textbf{MLP} & learning rate& $[1 \times 10^{-6},\ 0.9]$\\
& number of layers &$[1,\ 10]$\\
& number of neurons per layer& $[1,\ 200]$\\
& activity regularisation on each layer&$\{L1(\alpha=0.001)$, $L2(\alpha=0.001)$, None$\}$\\
\cmidrule(lr){2-2}\cmidrule(lr){3-3}
\textbf{CNN} & learning rate& $[1 \times 10^{-6},\ 0.9]$\\
& number of filters &$[1,\ 100]$\\
& number of dense layers &$[1,\ 10]$\\
& number of neurons per dense layer & $[10,\ 200]$\\
& activity regularisation on each layer&$\{L1(\alpha=0.001)$, $L2(\alpha=0.001)$, None$\}$\\
\cmidrule(lr){2-2}\cmidrule(lr){3-3}
\textbf{LSTM}& learning rate& $[1 \times 10^{-6},\ 0.9]$\\
& number of LSTM layers &$[1,\ 3]$\\
& number of LSTM units per layer& $[1,\ 100]$\\
& activity regularisation on each layer&$\{L1(\alpha=0.001)$, $L2(\alpha=0.001)$, None$\}$\\
\bottomrule 
\end{tabular}}
    \caption{The search domain for hyperparameters of the ML algorithms. For the MLP, CNN and the LSTM, the number of (dense) layers excludes the final densely-connected output layer with three units.}
    \label{table HPT
    search space}
\end{table}

\begin{table}[htbp!]
    \centering
    \resizebox{\textwidth}{!}{%
\begin{tabular}{lllllll} 
\toprule 
& & &  \multicolumn{4}{c}{Optimised value:}\\
\cmidrule(lr){4-7}
&&&\multicolumn{2}{c}{\ross}&\multicolumn{2}{c}{\lstt}\\
Algo&Target LLEs& Hyperparameter&1 time step & 6 time steps&1 time step & 6 time steps\\
\cmidrule(lr){1-1}\cmidrule(lr){2-2}\cmidrule(lr){3-3}\cmidrule(lr){4-4} \cmidrule(lr){5-5}\cmidrule(lr){6-6}\cmidrule(lr){7-7}
 \textbf{RT}&LLE 1&maximum depth&78&58&100&100\\
 &  &minimum examples per leaf&20&13&15&16\\
\cmidrule(lr){2-2}\cmidrule(lr){3-3}\cmidrule(lr){4-4}\cmidrule(lr){5-5}\cmidrule(lr){6-6}\cmidrule(lr){7-7}
 &LLE 2&maximum depth&100&24&100&78\\
 &  &minimum examples per leaf&20&20&1&1\\
\cmidrule(lr){2-2}\cmidrule(lr){3-3}\cmidrule(lr){4-4}\cmidrule(lr){5-5}\cmidrule(lr){6-6}\cmidrule(lr){7-7}
 &LLE 3&maximum depth&17&48&100&72\\
 &  &minimum examples per leaf&2&1&14&20\\
\cmidrule(lr){2-2}\cmidrule(lr){3-3}\cmidrule(lr){4-4}\cmidrule(lr){5-5}\cmidrule(lr){6-6}\cmidrule(lr){7-7}
\textbf{MLP}&all&learning rate&7.300e-05&1.222e-04&7.628e-05&1.939e-05\\
&  &number of dense layers&7&6&8&10\\
&  &number of neurons per dense layer&200&182&165&200\\
 &  &activity regularisation on each layer&L2&L2&L2&L1\\
\cmidrule(lr){2-2}\cmidrule(lr){3-3}\cmidrule(lr){4-4}\cmidrule(lr){5-5}\cmidrule(lr){6-6}\cmidrule(lr){7-7}
\textbf{CNN}&all&learning rate&1.081e-04&1.586e-04&3.287e-04&1.016e-04\\
&  &number of filters&100&37&29&59\\
 &  &number of dense layers&6&6&3&6\\
 &  &number of neurons per dense layer&200&200&122&200\\
 &  &activity regularisation on each layer&None&L1&L1&L1\\
\cmidrule(lr){2-2}\cmidrule(lr){3-3}\cmidrule(lr){4-4}\cmidrule(lr){5-5}\cmidrule(lr){6-6}\cmidrule(lr){7-7}
\textbf{LSTM}&all&learning rate&2.857e-03&1.874e-03&2.722e-03&1.105e-04\\
 &  &number of LSTM layers&3&1&2&2\\
 &  &number of LSTM units per layer&62&100&100&100\\
 &  &activity regularisation on each layer&L1&L1&L1&None\\
\bottomrule 
\end{tabular}}
    \caption{The optimal hyperparameter values selected by 50 iterations of Bayesian optimisation.  For all RTs, the optimal maximum leaf nodes was $100$, and the optimal minimum weight fraction per leaf was $0$. Some RT hyperparameters are excluded for brevity.}
    \label{table HPT results}
\end{table}

 \tab\ref{table HPT results} shows the optimised hyperparameter values, resulting from $50$ iterations of the optimisation
algorithm.
Some optimal
values are at the boundary of the search domain (\tab\ref{table HPT
search space}). 
This suggests that the algorithms might have performed better
with hyperparameter values beyond the chosen search domain. 
For instance, in every experiment configuration,
the optimised value of the RT hyperparameter \textit{maximum leaf
nodes} is $100$, the maximum value in the search domain. Other examples of
hyperparameters whose optimised values are at the boundary of the search domain for some configurations are the number of
neurons per layer for the MLP and the CNN, and the number of LSTM units per
layer. 
Nevertheless, the number of dense layers in the MLP and the CNN, as well as the
number of LSTM layers, are mostly not at the boundary. 
Overall, from the results of the hyperparameter optimisation, we can argue that although greater prediction accuracy might be achieved using NNs with larger layers (more neurons or LSTM units), the number of layers (NN depth) in the experiments is adequate.

To give an idea of the complexity of each of the four optimised ML algorithms, \tab\ref{table algo size} shows the size of the ML models, measured by the
maximum number of comparisons for the RT and the number
of trainable parameters in the case of the NNs.
The model size is a function of the optimised hyperparameter values.
For the NNs with $6$ time steps as input, the model size of those that predict the \lst LLEs
is larger than those that predict the \ros. 
This is perhaps reflective of the more chaotic dynamics of the \lst, which has a larger first LE as well as two non-linear terms in the
equations as opposed to only one in the \ros (see \eqs\ref{eqn Rossler}~and~\ref{eqn Lorenz}).
The same is not true of the NNs that take $1$ time step as input. 
This is unsurprising since the $1$ time step input contains less information on the
dynamics, \ie, how the state variables are changing in time.
In this case, the size of the MLP and the CNN is smaller in
the \lst, whereas the size of the LSTM is larger in the \lst. 
Due to their
architectural similarity in the $1$ time step case, it is unsurprising
that the optimal model size for MLP and CNN behaves similarly.

\begin{table}[htbp!]
    \centering
    \resizebox{0.8\textwidth}{!}{%
\begin{tabular}{rlllll} 
\toprule 
&&\multicolumn{2}{c}{\ross}&\multicolumn{2}{c}{\lstt}\\
Algo& &$1$ time step & $6$ time steps&$1$ time step & $6$ time steps\\
\cmidrule(lr){1-1}\cmidrule(lr){2-2}\cmidrule(lr){3-3}\cmidrule(lr){4-4} \cmidrule(lr){5-5}\cmidrule(lr){6-6}
 \textbf{RT}& maximum comparisons &99&99&99&99\\
\textbf{MLP}&trainable parameters& 242,603 & 170,537& 192,888   &  366,203 \\
\textbf{CNN}&trainable parameters& 222,503 &224,262 &  34,244  & 237,616  \\
\textbf{LSTM}&trainable parameters&78,557 &41,903 &  122,303  & 122,303\\
\bottomrule 
\end{tabular}}
    \caption{The size of the ML models, measured by the number of comparisons (RT) and number of trainable parameters (NNs). The maximum number of comparisons for each RT is $99$ (one less than the maximum leaf nodes), since the max depth is too large to constrain the number of non-leaf nodes in the tree.}
    \label{table algo size}
\end{table}

The maximum size of the RTs is the same for all configurations.
This is due to the tight restriction placed on the tree size by the
search domain of the maximum leaf nodes hyperparameter (\tab\ref{table HPT search space}).
However, the maximum depth hyperparameter varies across configurations and between
LLEs. 
The optimal value of the maximum depth depends on the number of linear separations of the input space
that
improves the predictions during the generation of the tree from the training data.
For the same input type, and for a given LLE, the RT for the \lst has a greater
maximum depth than the RT for the \ros.
The greater maximum depth implies that more accurate predictions can be made by
separating the input space at smaller scales (\ie making finer-grained partitions of the input space) in the \lst, compared to the \ros.
In other words, there is clearer detail at smaller relative scales (in the input space) in the \lst
than in the \ros.

\subsection{Results}\label{Section results subsection}
\paragraph{ Impact of data set size}
\begin{figure}[htb!]
    \centering
    \includegraphics[width=0.58\linewidth,  ]{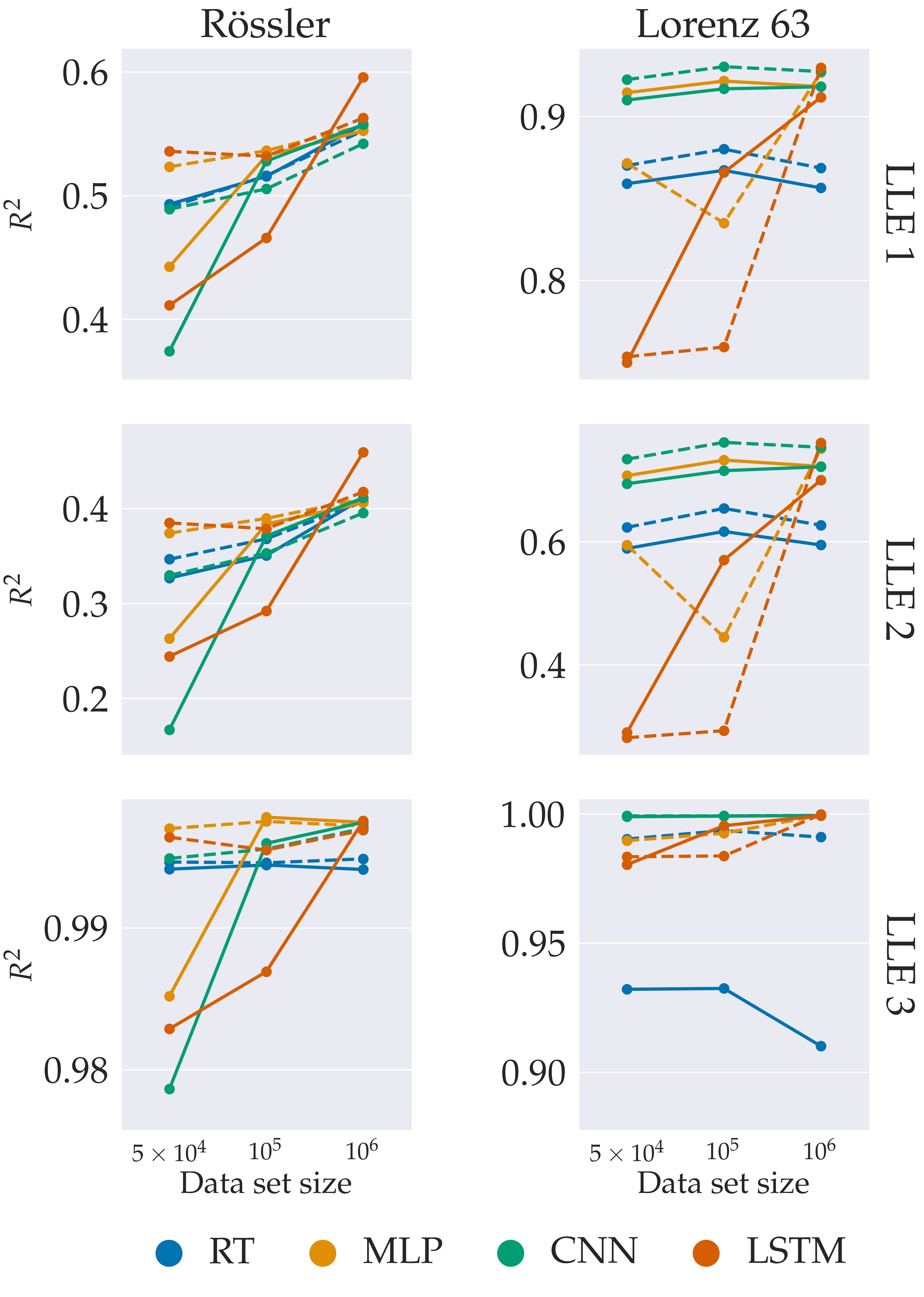}
    \caption{ The impact of data set size on mean \rs scores,
    across $30$ trials. The
    solid lines indicate $1$-time step results, the  dashed lines indicate
    $6$-time step results.}
    \label{fig data set size R2 means impact}
    \end{figure}

 As discussed in \s\ref{sec experiment setup}, whilst we focus on
results using a data set of $10^5$ examples, we ran all
experiments with three data set sizes to determine the impact on prediction
accuracy. We use the same setup for all data set sizes:
hyperparameters as described in \s\ref{section HPO}, $30$ trials, and a
partition of 
$0.6:0.2:0.2$ training-validation-testing. The mean \rs scores from all data set sizes are
shown in \fig\ref{fig data set size R2 means impact}. In both systems, there are only
small (or negligible) gains
in accuracy from using $10^6$ examples compared to $10^5$
examples. The one exception is the LSTM: in both systems, the \rs scores of the
LSTM
significantly increased with data set size. In the \ros, the LSTM becomes the most
accurate method when trained on $10^6$ examples, whilst in the \lst the
LSTM achieves accuracy similar to the MLP and CNN (indeed, \fig\ref{fig data set
size R2 means impact} shows the $6$-time step
input LSTM achieves the best mean \rs scores). In both systems, the
variation of the LSTM over the 30 trials is substantially reduced with $10^6$
examples 
compared to $ 10^5$ examples (notably, for the $1$-time step LSTM in \lstt, the variance of \rs
scores of LLE 2 reduces from $0.0201$ to $0.0007$). This suggests that choosing $10^5$
examples strongly limits the performance of the LSTM. We suspect that the LSTM
requires more data than the MLP and CNN due to its more complicated
architecture, namely the hidden state and the three parameterised gates which control
information flow into and out of the hidden state.

With regards to the
smaller data set of $5\times10^4$, \fig\ref{fig data set
size R2 means impact} shows that impact on accuracy is different in the two
dynamical systems: in the \ros, the accuracy of MLP and CNN (especially with the
$1$-time step input) is strongly reduced (compared to the $10^5$ data set),
whereas for \lstt the equivalent reduction in accuracy is small. This is likely due
to the greater sparsity of data points in the loop of the \ros. With these insights as context,
the remainder of 
\s\ref{Section results subsection} refers to results from the $10^5$ data sets,
unless stated otherwise.

\begin{figure}[htb!]
    \centering
    \textbf{$R^2$ scores of test data sets for all 30 data set instances}\par\medskip
    \includegraphics[width=0.58\linewidth,  ]{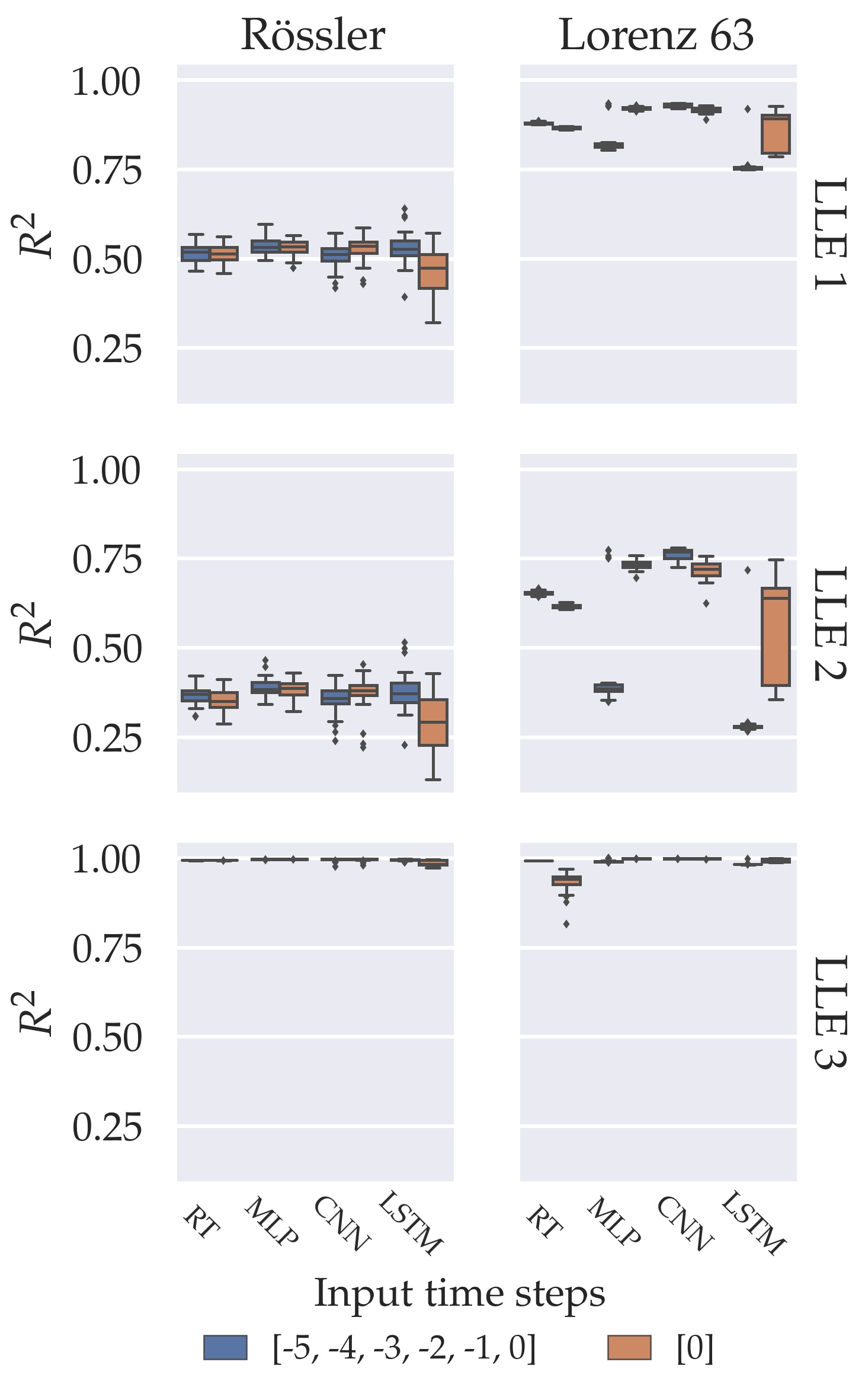}
    \caption{The $R^2$ scores of test data sets from the $30$ trials, showing the variation across data set instances, for each combination of system and ML method. Perfect predictions have an \rs score of one.  These results use the $10^5$ data sets:} each test data set has $20,000$ examples.
    \label{fig r2 test boxplots}
    \end{figure}

\paragraph{Comparisons between systems and across the LLE spectrum}
We assess accuracy with the \rs score of predictions made on the test data sets, each of which has
$20,000$ examples (see~\tab\ref{table data setup}). 
There are $30$ \rs scores for each configuration: one from each data set instance. 
These \rs scores are shown in boxplots in \fig\ref{fig r2 test boxplots} and summarised by their mean and standard deviation in
\tab\ref{table R2 score}. 
The immediate observation is that the \rs scores
differ consistently among LLEs (for a given system) and between systems (for a
given LLE).
The first LLE is predicted at least reasonably well in both dynamical systems ($0.54$ for \ros and $0.93$ for \lst).
The third LLE is well predicted in both systems, by all ML
algorithms, and with both types of input. 
Apart from one case, the mean \rs scores for LLE $3$ are
above $0.98$. 
This is unsurprising given the local homogeneity of LLE $3$ on the attractors, as
discussed in \s\ref{section system characteristics}.
In all cases, the second LLE is the least well predicted ($0.39$ for \ros and $0.76$ for \lst).
This result is to be expected since it is known that the second LLE is calculated least
accurately by the numerical method \citep{Kuptsov2012} and has a slower
convergence \citep{Bocquet2017}. Also, this result aligns with
some recent attempts to emulate chaotic dynamics with ML methods, where the emulators
have failed to reproduce the near-neutral LEs accurately \citep{Pathak2017, brajard20}.
We note that, particularly in
multiscale systems such as ocean-atmosphere systems, the neutral and
near-neutral exponents play an important role in understanding predictability,
(see \eg \cite{DeCruz2018, Quinn2020}) and are connected to the coupling
mechanisms \parencite{Vannitsem2016,Tondeur2020}.

Next we compare prediction accuracy between dynamical systems.
For LLEs
$1$ and $2$, predictions of the \lst tend to be better than those of the \ros. 
 The
highest mean \rs score for LLEs $1$ and $2$ are $0.9304$ and $0.7613$
(respectively) for \lstt, yet only $0.5365$ and $0.3897$ for \ross.
For LLE $3$, the mean \rs scores are similarly high in both systems.
These results indicate that the LLEs can be predicted and the variability of the prediction accuracy depends on which LLE and which dynamical
system is being predicted.

\begin{table}[htbp!]
    \centering
    \resizebox{\textwidth}{!}{%
\begin{tabular}{rcccccc}
\multicolumn{7}{c}{\Large \textbf{\rs on test data: Mean and (standard deviation) of 30 data set instances}}\\
\toprule
&\multicolumn{6}{c}{{\large \textbf{R\"ossler}}} \\
target & \multicolumn{2}{c}{LLE 1} & \multicolumn{2}{c}{LLE 2}&\multicolumn{2}{c}{LLE 3} \\
input type & 1 time step & 6 time steps & 1 time step & 6 time steps & 1 time step & 6 time
steps\\
\cmidrule(lr){2-2}\cmidrule(lr){3-3}\cmidrule(lr){4-4}\cmidrule(lr){5-5}\cmidrule(lr){6-6}\cmidrule(lr){7-7}
RT & 0.5155 (0.0248) & 0.5161 (0.0268) & 0.3506 (0.0299) & 0.3681 (0.0278) & 0.9944 (0.0002) & 0.9946 (0.0002) \\
MLP & 0.5323 (0.0211) & {\bf0.5363} (0.0243) & 0.3837 (0.0249) & {\bf0.3897} (0.0274) & {\bf0.9978} (0.0005) & 0.9975 (0.0006) \\
CNN & 0.5279 (0.0333) & 0.5054 (0.0349) & 0.3711 (0.0518) & 0.3530 (0.0419) & 0.9960 (0.0040) & 0.9956 (0.0044) \\
LSTM & 0.4657 (0.0633) & 0.5319 (0.0462) & 0.2921 (0.0769) & 0.3788 (0.0571) & 0.9869 (0.0074) & 0.9955 (0.0023) \\
\midrule
&\multicolumn{6}{c}{{\large \textbf{Lorenz 63}}} \\
target & \multicolumn{2}{c}{LLE 1} & \multicolumn{2}{c}{LLE 2}&\multicolumn{2}{c}{LLE 3} \\
input type & 1 time step & 6 time steps & 1 time step & 6 time steps & 1 time step & 6 time
steps\\
\cmidrule(lr){2-2}\cmidrule(lr){3-3}\cmidrule(lr){4-4}\cmidrule(lr){5-5}\cmidrule(lr){6-6}\cmidrule(lr){7-7}
RT & 0.8672 (0.0025) & 0.8801 (0.0027) & 0.6166 (0.0046) & 0.6540 (0.0053) & 0.9324 (0.0305) & 0.9936 (0.0003) \\
MLP & 0.9217 (0.0038) & 0.8350 (0.0438) & 0.7325 (0.0123) & 0.4449 (0.1446) & {\bf0.9993} (0.0003) & 0.9925 (0.0033) \\
CNN & 0.9169 (0.0081) & {\bf0.9304} (0.0047) & 0.7153 (0.0261) & {\bf0.7613} (0.0159) & 0.9992 (0.0005) & {\bf0.9993} (0.0002) \\
LSTM & 0.8659 (0.0530) & 0.7594 (0.0302) & 0.5702 (0.1419) & 0.2933 (0.0803) & 0.9955 (0.0044) & 0.9838 (0.0028) \\
\bottomrule 
    \end{tabular}}
    \caption{ The table shows mean \rs scores over 30 trials,
    and the corresponding standard deviations in parentheses. The \rs score
    measures the accuracy of predictions: the optimum score is 1. We calculate
    the \rs on the test data set for each of the 30 trials. The highest mean \rs
    score for each combination of LLE and system (for both input types) is shown
    in bold. ``1 (6) time step(s)'' refers to the number of time steps in the
    input.}
    \label{table R2 score}
\end{table}

\begin{figure}[htb!]
    \centering
    \includegraphics[width=0.8\linewidth,  ]{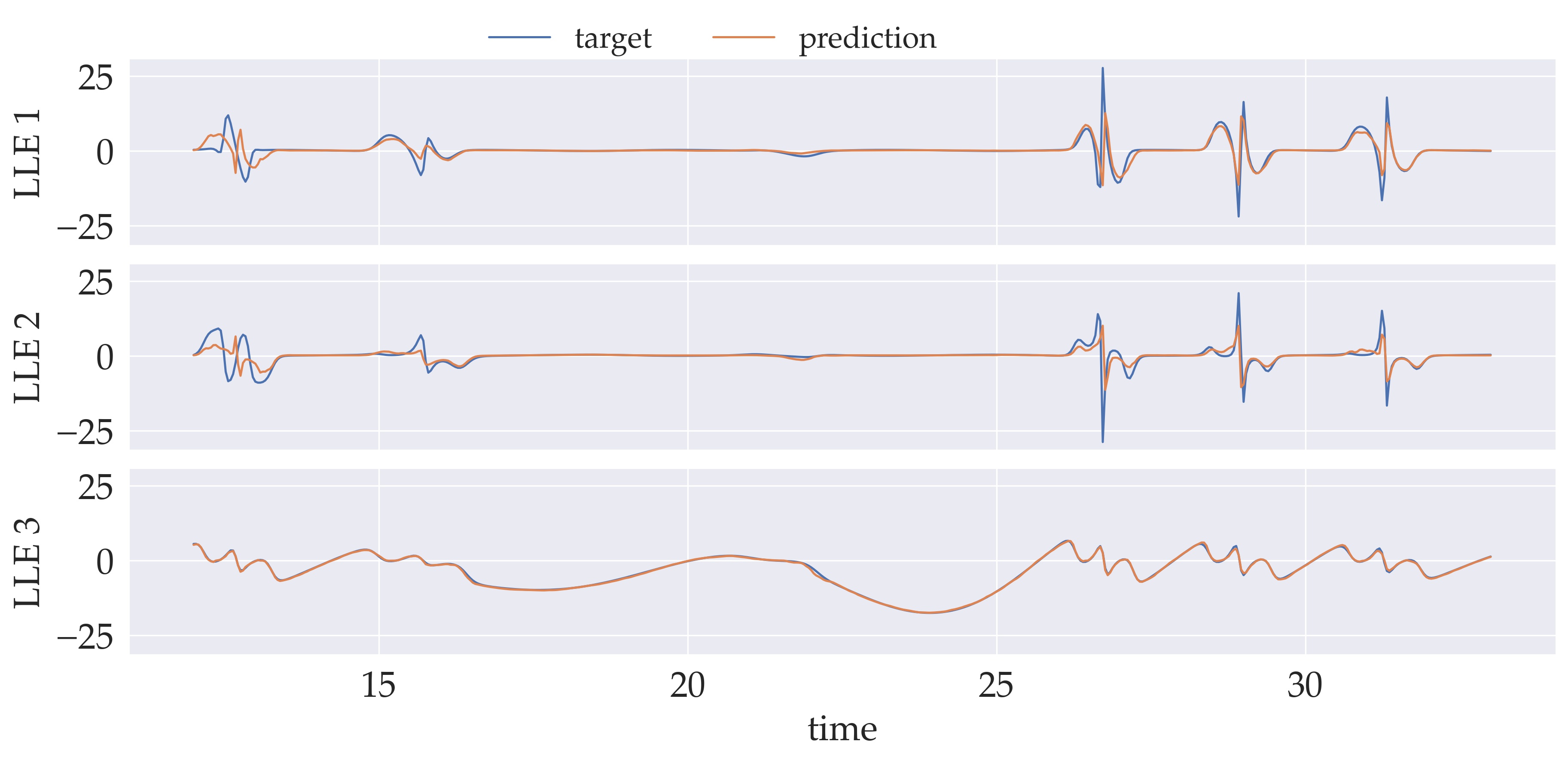}
    \caption{Time series of targets and predictions of test data from the \ros. Predictions made by an MLP with 6 input time steps. The \rs scores for the period shown are 0.4393, 0.3175 and 0.9981 for LLEs 1, 2 and 3, respectively.} 
    \label{fig time series MLP Rössler}
    \end{figure}

\begin{figure}[htb!]
    \centering
    \includegraphics[width=0.8\linewidth,  ]{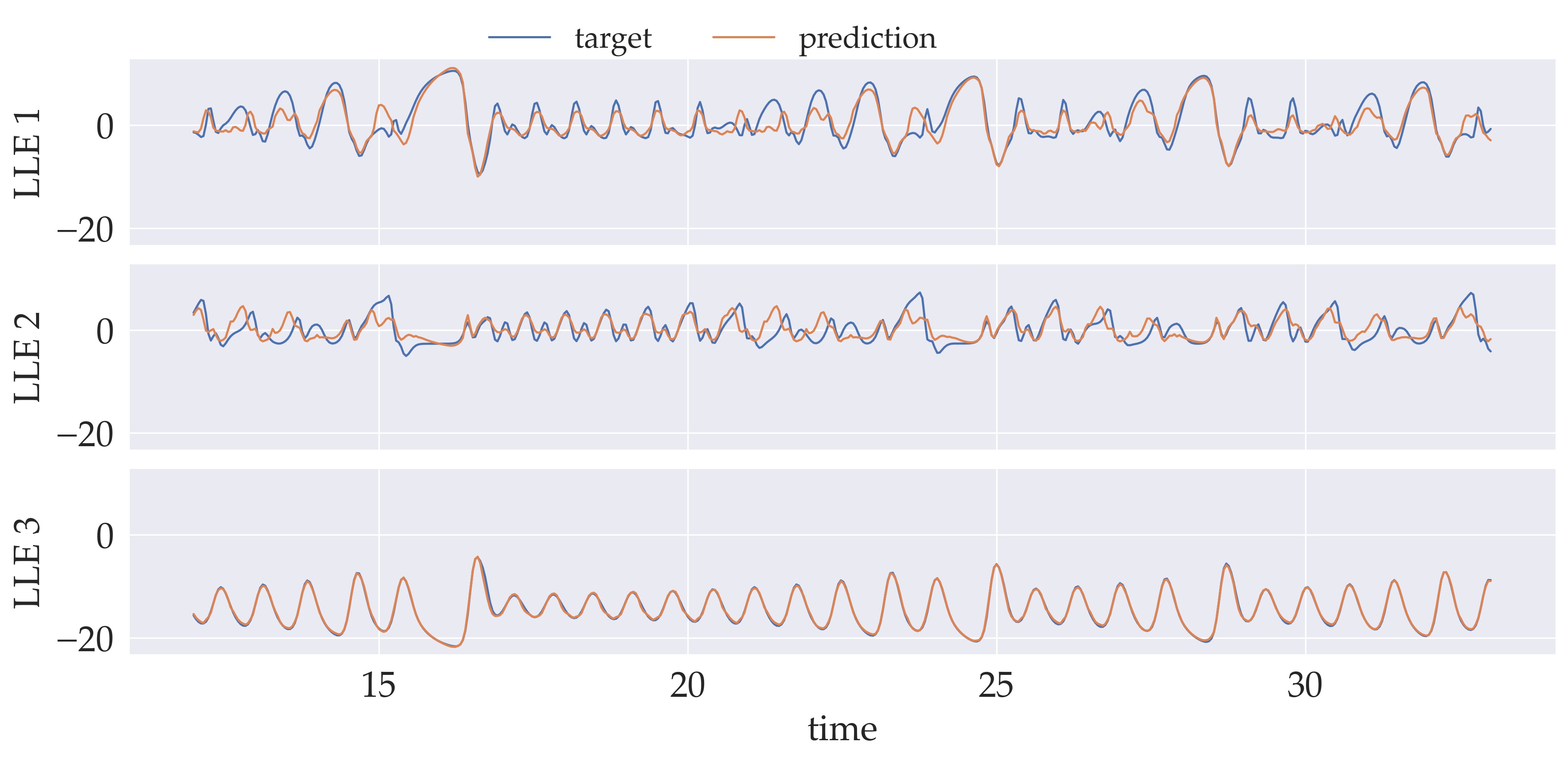}
    \caption{As with \fig\ref{fig time series MLP Rössler} for the \lst. The predictions are made by a CNN with 6 input time steps. The \rs scores for the period shown are 0.7981, 0.4897 and 0.9953 for LLEs 1, 2 and 3, respectively.} 
    \label{fig time series CNN L63}
    \end{figure}

\paragraph{Analysis of predictions on ordered test data} \figs\ref{fig time series MLP Rössler}~and~\ref{fig time series CNN L63} show
time series of target values and predictions for a small set of ordered test
data. 
The predictions are produced by the algorithm that achieves the best mean
\rs scores  (on the $10^5$ data sets): MLP for the \ros, and CNN for the \lst.
In both systems, LLE~3 is almost perfectly predicted throughout the time series.
However, the predictions of LLEs 1 and 2 have error characteristics that are specific to each system.

\fig\ref{fig time series MLP Rössler} illustrates that the first and second LLEs
of the \ros vary intermittently:
they are stationary and near the mean value for the majority of the time and then abruptly change and oscillate for a short period before returning to be close to the mean.
This corresponds to the system trajectory being in the disc in the $xy$-plane, and then jumping into the ``loop'' with positive $z$-values, before returning to the disc.
Predictions are extremely good in the stationary periods, and they are
satisfactory during the peaks, which we label ``fluctuation events''. This is
particularly true for LLE~1 where we see that the ML-based predictions always
catch the fluctuation event and often its sign. The predictions of
LLE~2 follows similar behaviour to LLE~1, however the \rs score suggests that the
pointwise accuracy is slightly worse than for LLE~1. 

On the other hand, in the \lst, \fig\ref{fig time series CNN L63} shows that LLEs 1 and 2 are constantly oscillating.
Certain characteristics of the target time series are well reproduced by the
predictions, \eg the largest peaks of LLE~1. 
These large peaks occur when the system trajectory passes close to the origin ({\it cf.} \fig\ref{fig distributions of LLEs Lorenz}), a region in which LLE~1 is locally homogeneous on the attractor.
Nonetheless, small errors occur frequently.
Notably, the higher-frequency features (such as the secondary peaks of LLE 2 between $t=17$ and $t=20$) are often relatively poorly reproduced for LLEs 1 and 2.

These time series provide further insight into the lower \rs scores for LLEs 1 and 2 in the \ros compared to the \lst.
Recall the definition of \rs in \eq\ref{eqn define R2 score}: the distance from the perfect score of $1$ is the sum of squared residuals divided by the total sum of squares. 
The periods of stationarity in the \ros LLEs 1 and 2 contribute little to the total sum of squares. Consequently the larger errors during fluctuation events strongly reduce the \rs score. 
In contrast, 
in the \lst, the \rs score is high despite more frequent prediction errors
because the constant variation of the target values results in a larger total
sum of squares. 

\paragraph{Impact of local heterogeneity in phase space}
The variability of the LLEs on the strange attractor of chaotic systems is a
known feature, the immediate consequence of which is a highly
state-dependent predictability horizon: two slightly different initial
conditions can generate trajectories with hugely different degree of
instability. In a recent work, \textcite{lucarini2020new} have for the first time
shown how this variability is related to the presence and distribution of
unstable periodic orbits, each with a different degree of instability, densely
filling the attractor. Arbitrary solutions are bounced among these unstable
periodic orbits taking their local instability features when they are in their
proximity. 

\begin{figure}[htbp!]
    \centering
    \begin{tabular}{cc}
\includegraphics[width=0.41\linewidth]{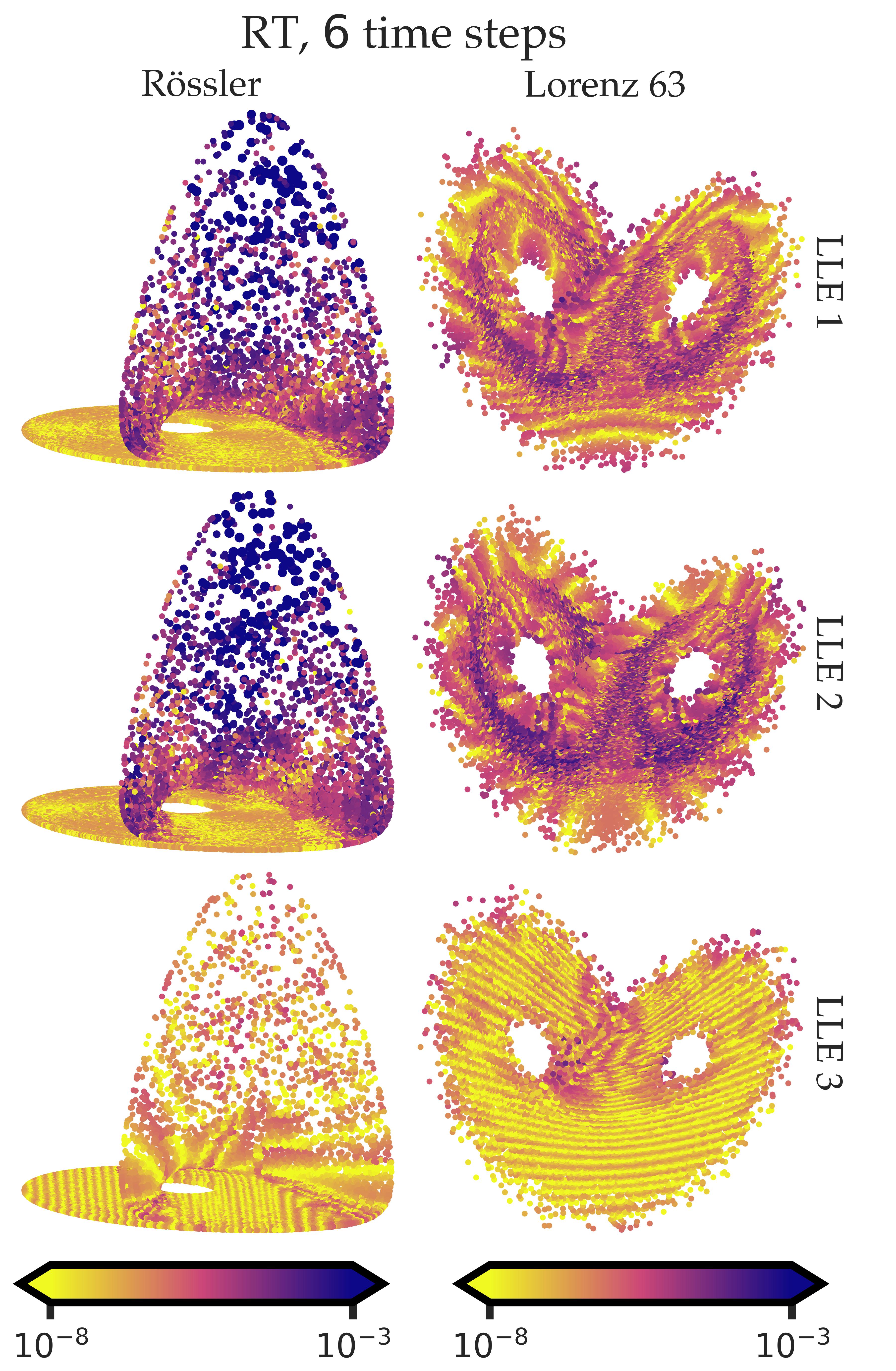}& 
\includegraphics[width=0.41\linewidth]{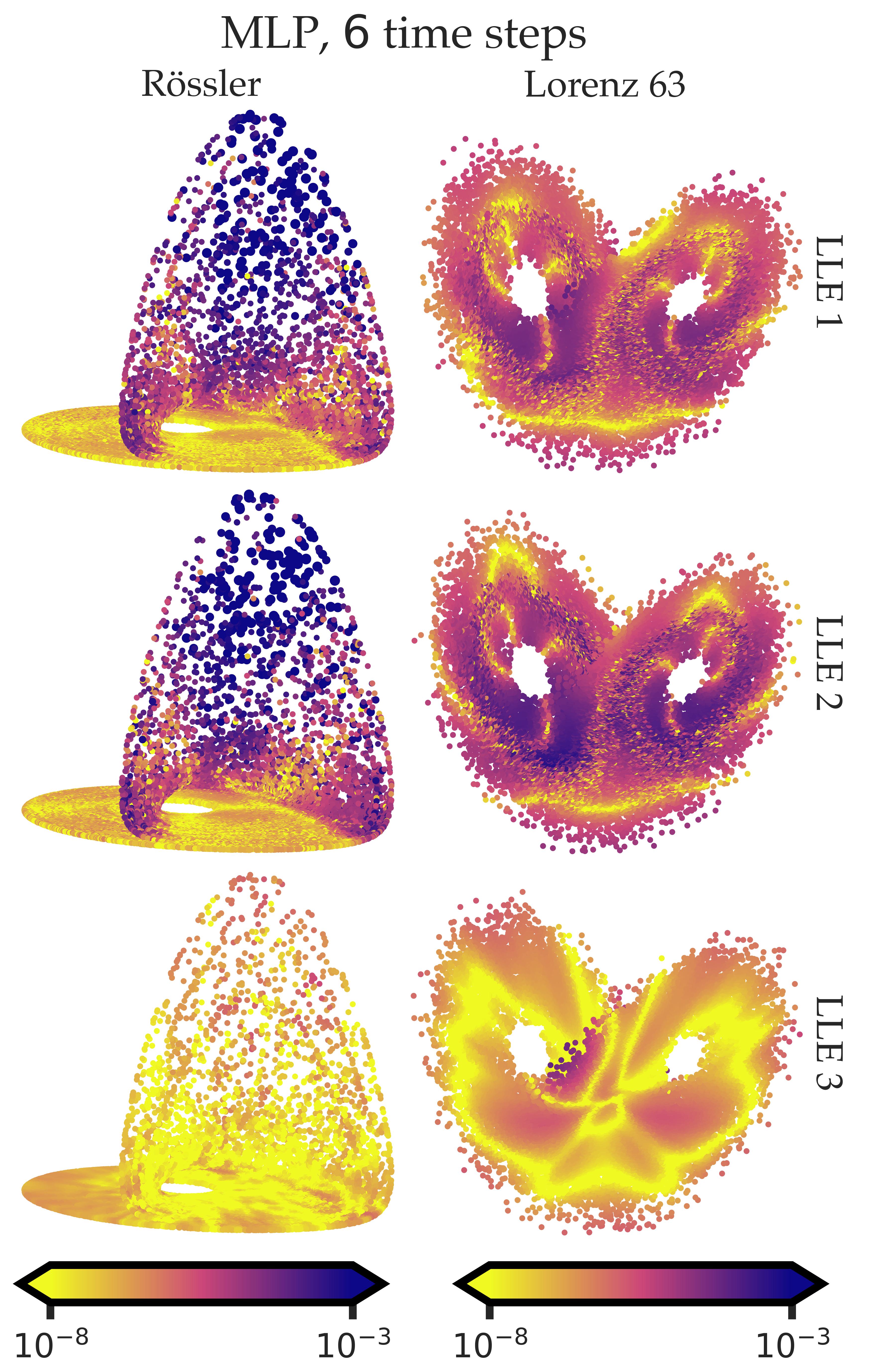} \\
\includegraphics[width=0.41\linewidth]{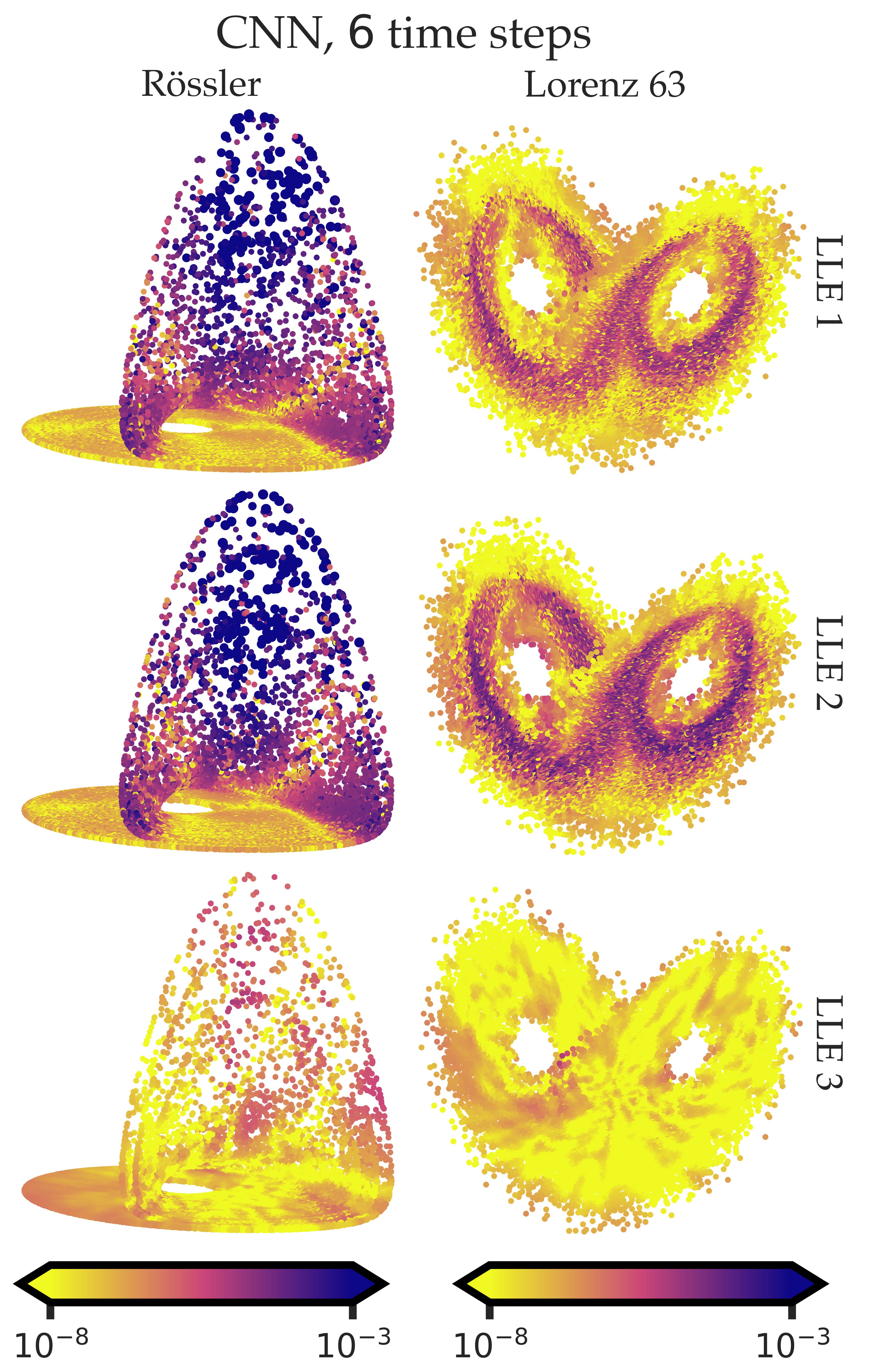} &
\includegraphics[width=0.41\linewidth]{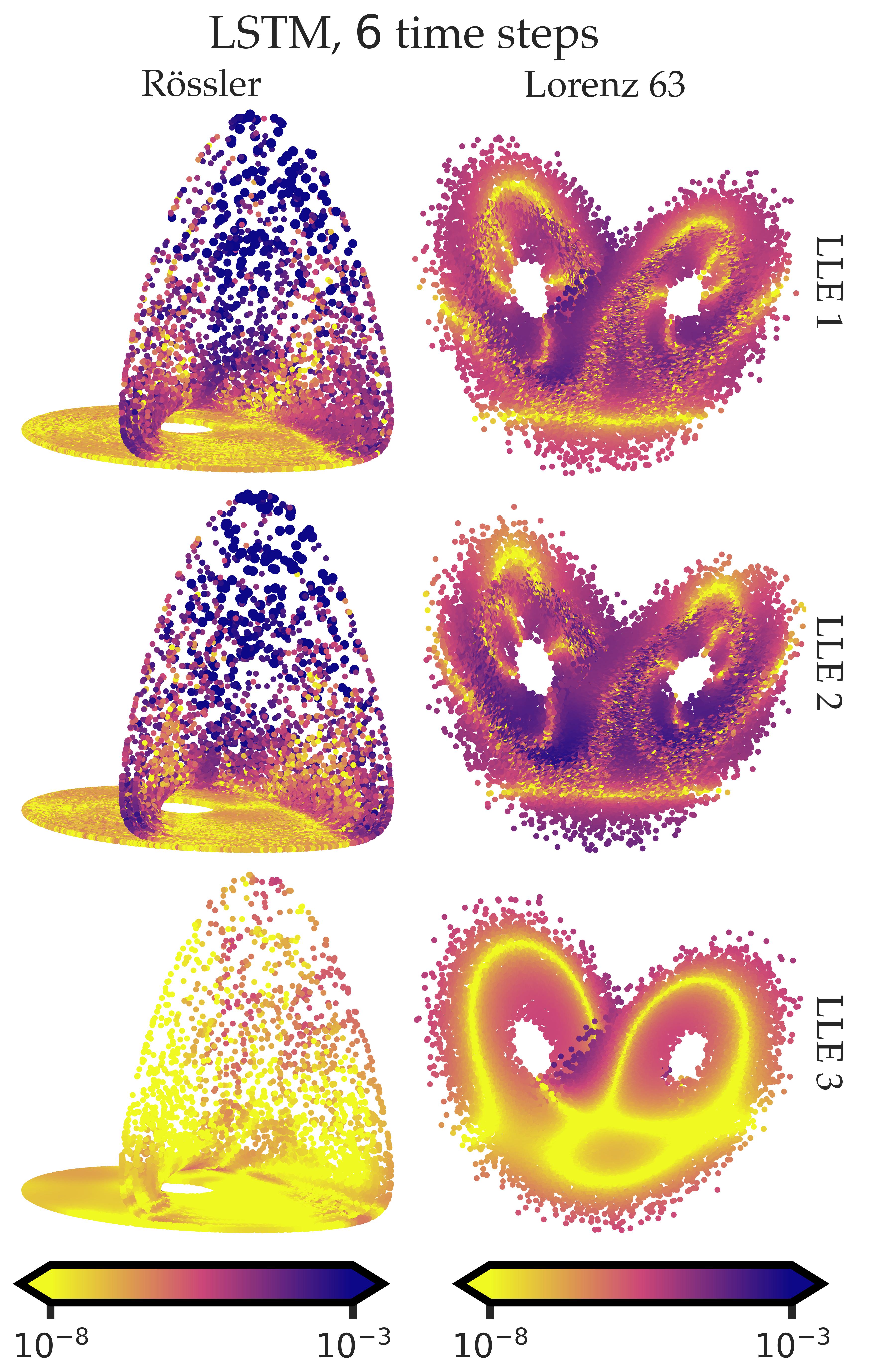}
    \end{tabular}
    \caption{Recall the definition of the \rs score (\eq\ref{eqn define R2
    score}): the distance from the perfect score of $1$ is given by the sum of
    squares of residuals divided by the total sum of squares. Here we give a
    local (in phase space) description of the contribution towards that distance
    made by each prediction in a set of $20,000$ test examples. Points are
    located at the current time step $\x_k$ of the input, and both coloured and
    sized by the square of the residual, divided by the total sum of squares.
    Darker points contribute a greater reduction to the \rs score. If we denote the colour values as $a_j$, then $R^2 = 1-\sum a_j$.
    We show all configurations with a $6$ time step input. For each configuration, we use the data set instance for which the \rs score was closest to the mean (as shown in \tab\ref{table R2 score}).}
    \label{fig 3d R2 contribution} 
\end{figure}

Recall from \s\ref{section system characteristics} that in both dynamical
systems, there are regions of the system's attractor where the values of LLEs~1
and 2 are
locally heterogeneous (LLE~3 is everywhere locally homogeneous). 
The locally heterogeneous regions in the \ros are in the loop with positive
$z$-values, and in the \lst they form a strip that lies halfway between the
outside edge and the centre of each wing. 
\fig\ref{fig 3d R2 contribution} shows where the larger prediction errors occur
on the attractor, for all configurations with a $6$ time step input.
More precisely, it shows the detraction from the perfect \rs score of 1 contributed by each point.
We see that for all ML algorithms, larger errors occur in the locally heterogeneous regions.
Moreover, the locally heterogeneous regions are robustly difficult:
for the most accurate algorithm in the \lst (CNN),
larger prediction errors only occur in these regions.
This suggests that local heterogeneity plays a key role in determining where on
the attractor it is possible for ML to make reliably accurate predictions of LLEs.

\begin{figure}[htbp!]
    \centering
    \begin{tabular}{cc}
\includegraphics[width=0.41\linewidth]{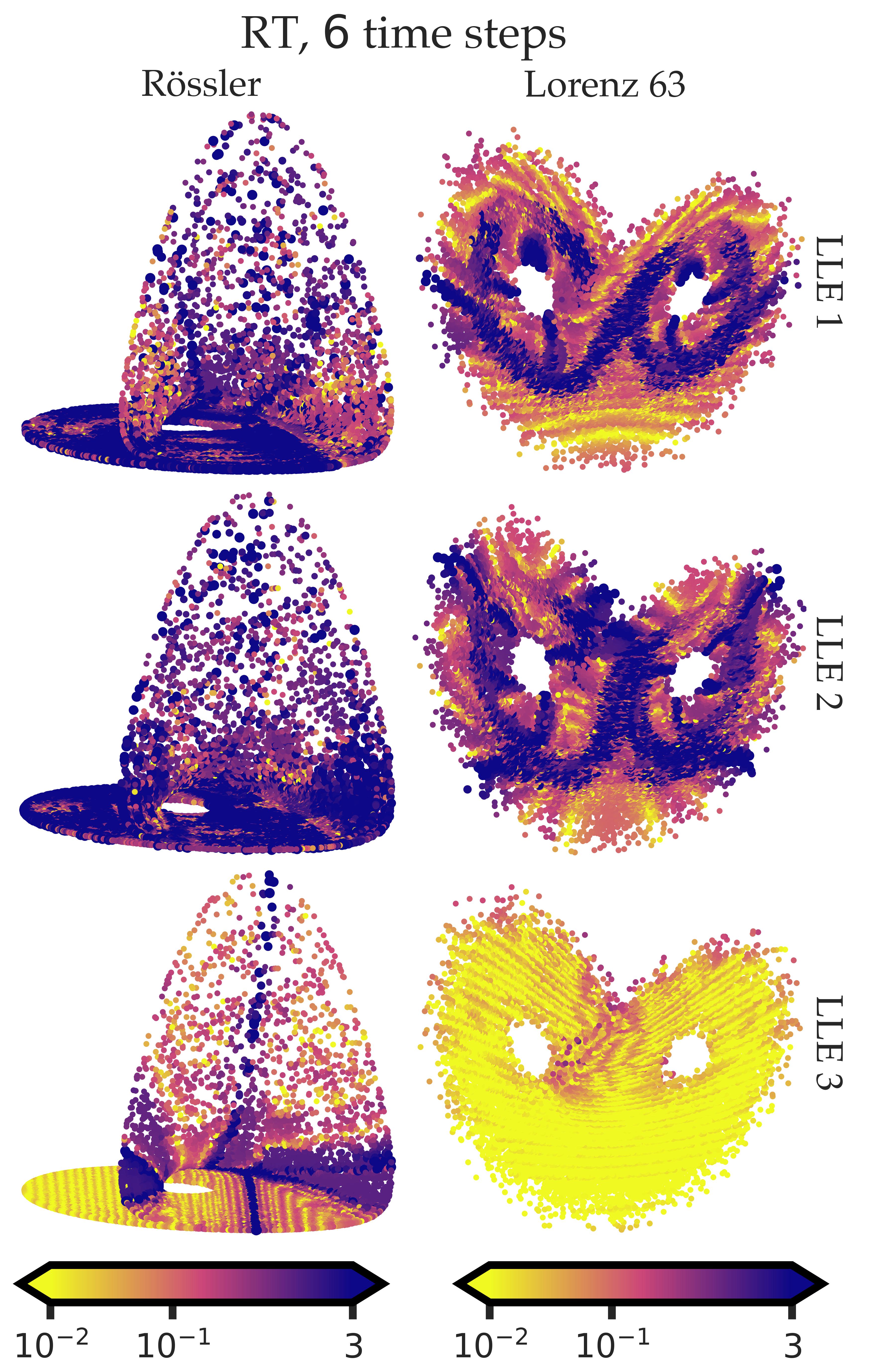} & 
\includegraphics[width=0.41\linewidth]{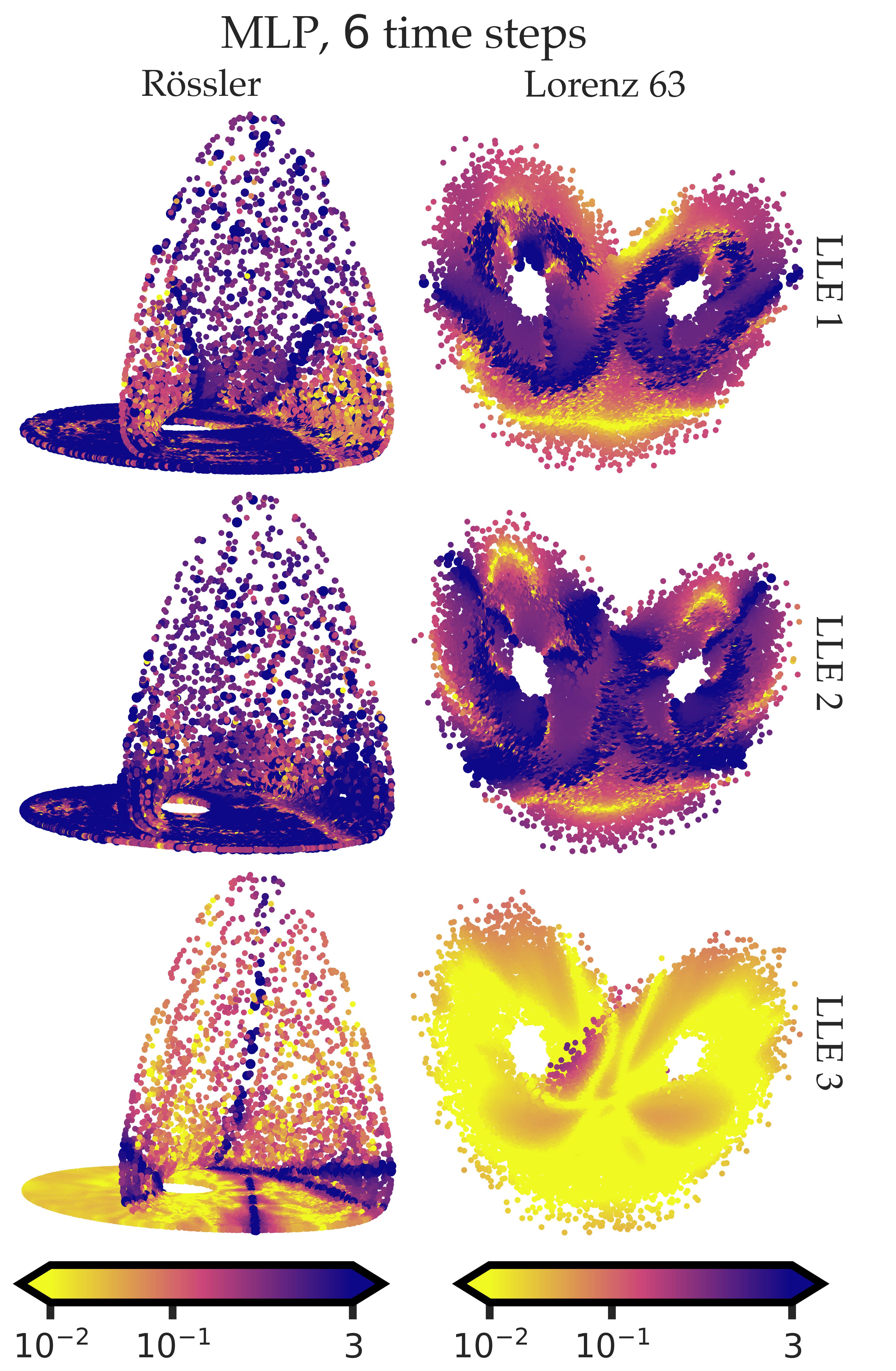}\\
\includegraphics[width=0.41\linewidth]{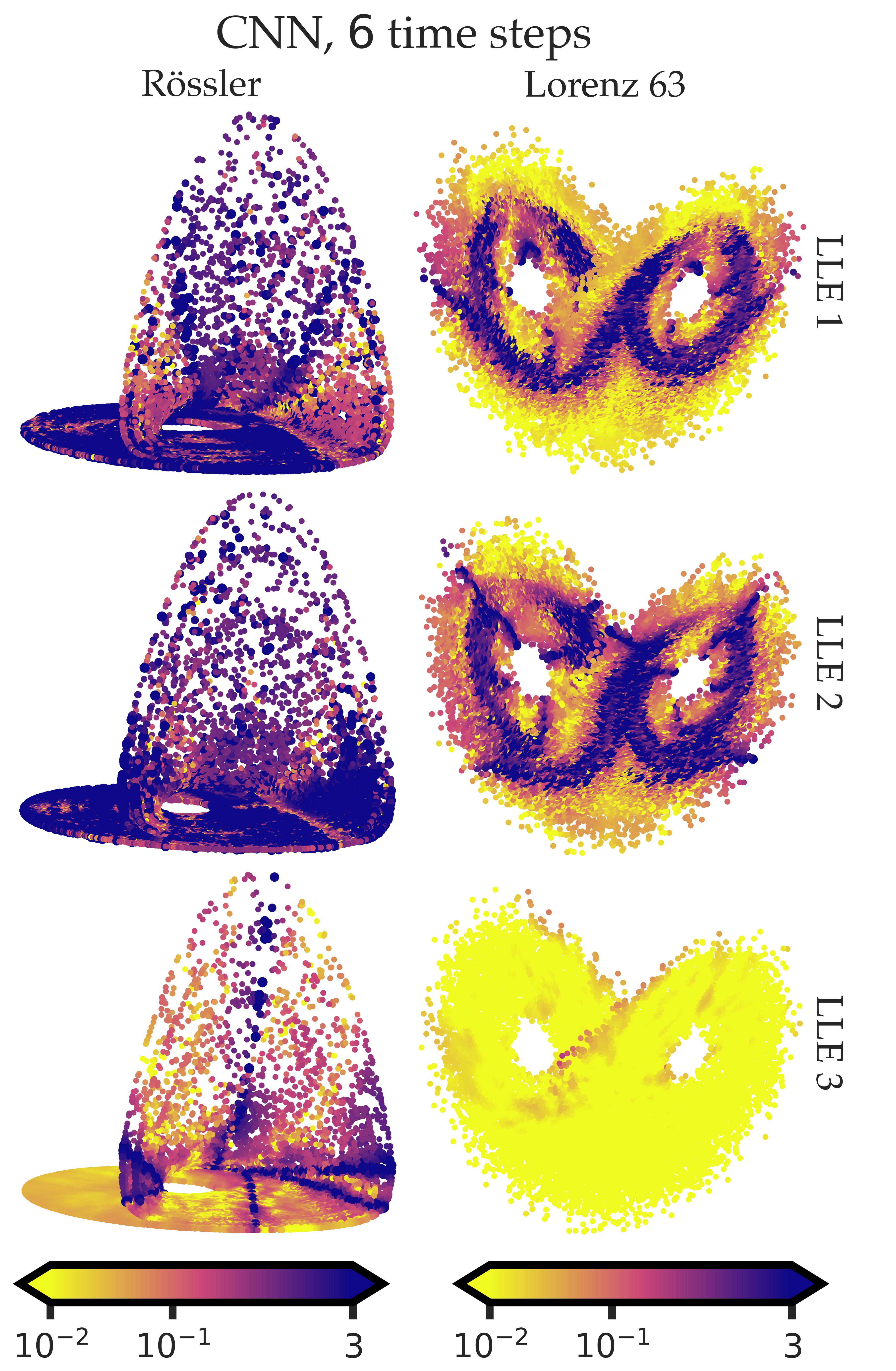} &
\includegraphics[width=0.41\linewidth]{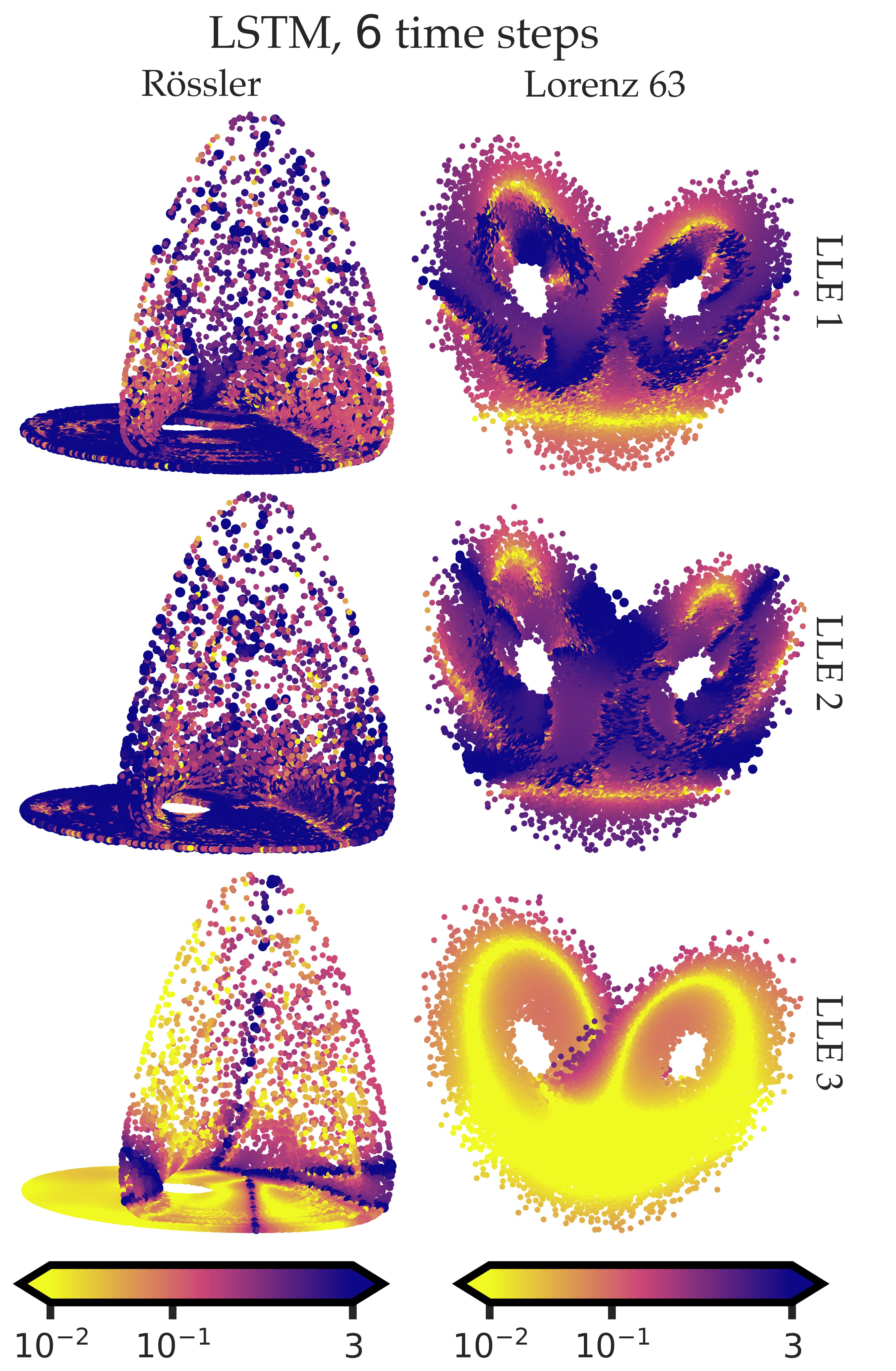}
    \end{tabular}
    \caption{As with \fig\ref{fig 3d R2 contribution}, but point colour and size
    show the absolute relative error, given by $|y-\hat{y}|/(\epsilon +
    |\hat{y}|)$, where $\epsilon = 10^{-6}$, $y$ is the prediction and $\hat{y}$
    is the target. 
    The darker the point, the larger the absolute relative error of the prediction.
    }
    \label{fig 3d relative residual}
\end{figure}

\fig\ref{fig 3d relative residual} shows that a similar pattern occurs for the absolute relative error of predictions. The relative error compares the size of the error to the size of the target value. Notably, there are some large relative errors in the locally homogeneous regions of the \ros attractor (\ie in the disc) since the relative error is especially punitive when the target value is close to zero.

The difficulty of making accurate predictions in these regions is in line with
ML theory.
ML algorithms work by optimising a model (such as a NN) to approximate the
map from the input to the target of the training data.
ML algorithms are successful if the optimised model also approximates the map
from input to target on unseen data, such as test data.
This is possible only if the
training data provide enough information about
the unseen data.
The fundamental problem in the locally heterogeneous regions is that the
training data cannot provide enough information because the LLE values are noisy.
In other words, the target values are highly variable even as length scales tend
to zero.
Thus the target values of unseen data are likely
to be quite different to those of nearby seen data.
Consequently, as local heterogeneity increases, ML models are less able to
generalise from training data to unseen data.

The characteristics of the local heterogeneity explains the differences in
prediction accuracy between the two dynamical systems. In the \ros, local
heterogeneity in the loop of the attractor results in poorer predictions
during the aforementioned fluctuation events. 
As explained above, errors during fluctuation events strongly reduce the
\rs score.
On the other hand, in the \lst the values of LLEs~1 and 2 in their respective
heterogeneous regions (see \fig\ref{fig distributions of LLEs Lorenz}) are
relatively close to the mean: the largest deviations from the mean are in locally homogeneous regions.
Therefore, the prediction errors resulting from locally heterogeneous regions
are likely to be small compared to the deviation of the target values from the
mean.
Consequently, the differences in local heterogeneity explain the higher \rs
scores achieved for LLEs~1 and 2 of the
\ros, compared to the \lst.

\paragraph{Impact of statistical distribution of targets and predictions}
The poorer prediction accuracy for the \ros is also explained, to a lesser
extent, by the statistical distribution of the LLE values. As described in \s\ref{section
system characteristics}, the \ros LLEs~1 and 2 include ``extreme events'', \ie values of large magnitude that appear infrequently. 
Predicting these extreme events is very challenging for any model, and
particularly so for ML: one would need to enlarge the training data set
commensurately to the (very long) return times of the extreme events.
 As noted above, we found a marked improvement in the LSTM
performance with a $10$-fold increase in data set size (see \fig\ref{fig data
set size R2 means impact}).

\begin{figure}[htb!]
    \centering
    \includegraphics[width=0.58\linewidth]{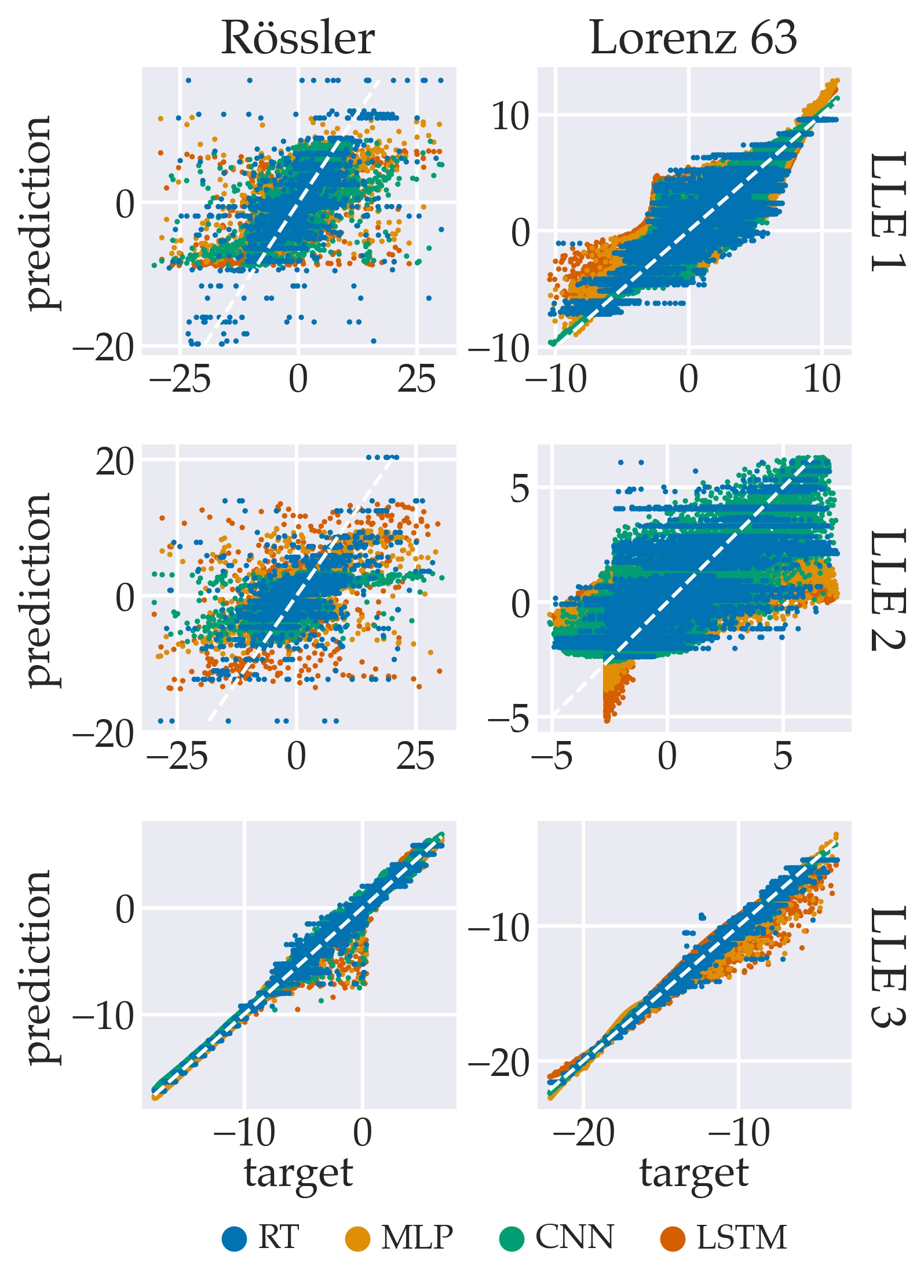}
    \caption{Scatter plots of targets vs. predictions, for a test data set. Note that the axis scales differ in each panel. For each method-system combination, the data are from the data set instance with \rs closest to the mean for all data set instances. Only results from setups with 6 input time steps are shown. Note that all scatter plots show different levels of heteroskedasticity, \ie the variance of a predicted value depends on the value of the target. This is a well-known challenge for regression methods.}
    \label{fig scatter target predictions}
    \end{figure}

\begin{figure}[htb!]
\centering
\includegraphics[width=0.58\linewidth]{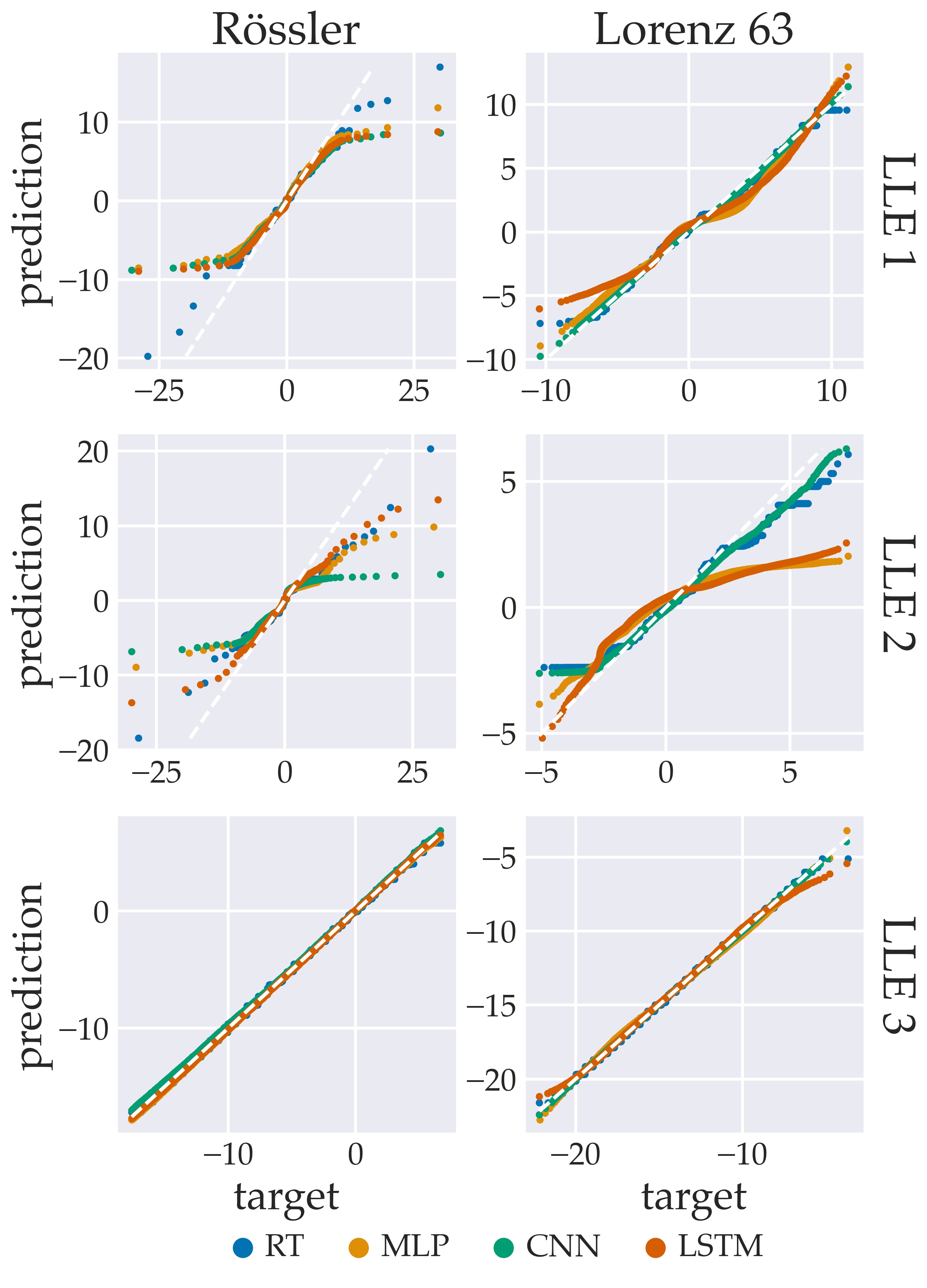}
\caption{The $1000$-quantiles of the predictions are plotted against those of the targets, revealing how well the two distributions match. The closer the graph to the $y=x$ line (dotted white), the closer the prediction distribution to the target distribution. Note that there is not necessarily any relationship between proximity of distributions and (point-wise) accuracy of predictions. In each panel, and for each machine learning method, the quantiles are from the test data of the data set instance with \rs closest to the mean. Only results from setups with 6 input time steps are shown.}
\label{fig qq plots}
\end{figure}

A complementary picture of the prediction accuracy is given in \fig\ref{fig scatter target predictions} that shows target values plotted against
predicted values.
The panels for the first and second LLEs of the \ros show that all four ML algorithms
fail to predict the larger magnitude targets accurately. Similarly, the QQ-plots in \fig\ref{fig qq plots} show that the predictions fail to
replicate the extremities of the true values: the minimum and maximum of the
predictions are lower in magnitude than those of the target.
For instance, for LLE~1, we see that the NN methods make few predictions greater than $10$
in magnitude, despite target values reaching magnitudes of $25$. 
This behaviour can also be seen in the time series plot (\fig\ref{fig time series MLP Rössler}).
In the \lst, however, the larger-magnitude targets occur more frequently and are thus well-represented in the training data set.
Consequently, we see in \fig\ref{fig qq plots} that the larger magnitudes are well
represented in the predictions distribution, even
if the predictions sometimes fail to capture the amplitude of the targets on a pointwise
basis, as illustrated in \fig\ref{fig time series CNN L63}.

\paragraph{Impact of ML approaches to exploiting temporal structure}
Each of the ML algorithms we test takes one of three approaches to exploiting the temporal structure of the
input, as discussed in \s\ref{section SL algos}.  A comparison of these
approaches can only be made when there is non-trivial temporal structure: this
paragraph refers only to results
with the $6$-time step input.
We find that there is no single optimal approach across both systems and all
data set sizes. \fig\ref{fig r2 test boxplots} shows that with the $10^5$ data set, the MLP and LSTM perform comparatively
well in the \ros but far more poorly in the \lst. However, with the $10^6$ data set, clearer patterns emerge: in
both systems, the LSTM is the most accurate, and the MLP and CNN perform
similarly well. With the larger data set, there is not a clear distinction in
performance between the three approaches.

 \paragraph{ Impact of input type}
  With the $10^5$ data set, the input type has a big impact on
the MLP and LSTM algorithms, particularly for LLE $2$ of \lstt: the mean \rs
 score for MLP is $65\%$
  better with $1$-time step than with $6$, and for the LSTM
 it is $94\%$
  better. However, with the $10^6$ data set, clearer patterns emerge:
 in the \ros, the $1$-time step input achieves the same or better mean~\rs for all algorithms,
 whereas in \lstt, the $6$-time step input achieves higher mean~\rs for all
 algorithms, and lower
 variance of \rs for all NN algorithms. This may be due to the more
 complicated dynamics of the \lst (see \s\ref{section system characteristics}),
 or due to \lstt having a smaller Lyapunov time ($1.1$ compared to $5.1$ in
 the \ros),  meaning that $6$ time steps provides more information in \lstt. As
 mentioned in \s\ref{sec experiment setup}, in an idealised case (without
 computational constraints) one would optimise the input to include more distant
 time steps. The comparison between the $10^5$ and $10^6$ data sets shows that
 some NN algorithms require more data in order to fully exploit the additional
 information provided in the $6$-time step input. With the $10^6$ data set, the
 maximal difference of mean \rs scores between input types (for any given algorithm) is small: $1$-time
 step is $10\%$ better
  in
 \ross (LSTM, LLE 2) and $6$-time steps is $9\%$ 
 in \lstt (LSTM, LLE 2).

 Among two input types we tested, the \rs scores
in \tab\ref{table R2 score} (for the $10^5$ data set) suggest that across all algorithms, there is only a small advantage to be
gained by providing $6$ time steps in the input rather than $1$. However, for
a given algorithm, the input type can have a big impact. \fig\ref{fig data set
size R2 means impact} shows this is especially true
for the smaller data set of $5\times 10^4$ in the \ros, where the NN's prediction
accuracy were far lower with the $1$-time step input.

\paragraph{Computation time}\label{sec computation time}
As stated previously, this is a feasibility study focused on accuracy. Nevertheless, we discuss here briefly the computational cost, keeping in mind however that optimising the latter was not our priority. These arguments are thus included for context and completeness, and not as a proof of viability.
\tab\ref{table computational cost} shows the mean (and standard deviation) time
elapsed per prediction of the
three LLEs, computed over 4000 trials on a CPU processor in a personal computer.
The RT is two
orders of magnitude faster than the NNs. This is expected: the
RT has far fewer trainable parameters (see \tab\ref{table algo size}) and, unlike the NNs, does not require the
evaluation of activation functions.
The RT is the only ML algorithm that is faster than the standard method of
computing LLEs (propagating perturbations and orthogonalising, ignoring the time
taken by the spin up iterations).
Despite the comparative simplicity of the RT, it achieves \rs scores that are
close to those of the NNs, although it is less close in the \lst.
It is likely that the number of output values (constrained by the
maximum leaf nodes hyperparameter) limited the RTs' performance more strongly
in the \lst, due to the greater complexity of the \lstt attractor.

\begin{table}[htbp!]
    \centering
    \resizebox{0.7\textwidth}{!}{%
\begin{tabular}{rcccccc}
\toprule \multicolumn{7}{c}{\textbf{Execution time per set of 3 LLEs}}\\
\midrule
& & Standard Method& RT& MLP&CNN&LSTM\\
{\textbf{R\"ossler}} & mean & 2.32e-4 & 9.96e-5&2.37e-2&2.38e-2&2.34e-2 \\
& $\sigma$ & 2.16e-5&3.51e-6&2.36e-2&2.57e-3&2.51e-2\\
{\textbf{Lorenz 63}} & mean & 9.69e-4 & 9.88e-5 & 2.36e-2 & 2.37e-2&2.31e-1\\
& $\sigma$ &2.53e-3&2.26e-6&2.85e-2&2.83e-2&1.60e-3\\
\bottomrule 
\end{tabular}}
    \caption{The time elapsed for one set of LLEs to be calculated (by the standard method) or predicted (by the ML algorithms). The table shows the mean and standard deviation of elapsed time from 4000 trials.}
    \label{table computational cost}
\end{table}

Finally, we speculate on the cost of making predictions in the
sub-problem (see \s\ref{subsec problem
statement and eval metrics}).
The energy cost (measured in flops) of the RT scales linearly with the number of LLEs, since a separate
tree is trained for each scalar target. However, the time cost remains the same
as the RTs can be executed in parallel. For the NNs, the potential cost saving
is not clear without further
experimentation. Less expensive NNs could be attained via pruning or distilling
methods, \eg see \textcite{Molchanov2017}.

\section{Discussion and summary}\label{section discussion}

This study discusses the use of supervised machine learning (ML) to support numerical forecasting of chaotic dynamics. A huge amount of work has appeared recently at the crossroads between ML and the geosciences, whereby the former has provided novel data-driven solutions to complement or, in some ideal scenarios, to substitute the physical models \cite[see \eg][]{Sonnewald2021}. In this work we took a different approach that we referred to as ``non intrusive''. 

We did not pursue improving the given physical model with a ML model, but rather using ML models as a supplementary tool that provides information to drive adaptive decisions while running the prediction. 
The range of possible desirable information is ample, \eg anticipating a regime change or the onset of intense convective events, as is that of consequent actions. In this work, we focused on chaotic systems where real time knowledge of the unstable properties of the system's state is of paramount relevance. Local Lyapunov exponents (LLEs) provide this knowledge in the form of the local (in time) exponential rates at which errors about the system's state evolve \citep{Benettin1980theory, Pikovsky2016}. Nevertheless, they are notably difficult to compute, require the coding and maintaining of a tangent linear model, and the computational cost grows fast with the system's size.
This work is a feasibility study that investigates the accuracy
with which supervised ML can estimate the LLEs of a dynamical system trajectory based only on the system state at the current time step and a few recent time steps.

 We tested four supervised ML algorithms (a regression tree (RT), a multilayer
 perceptron (MLP), a convolutional neural network (CNN) and a long short-term
 memory network (LSTM)) on two dynamical systems (the \ros and \lstt systems).
 The dynamical systems are chaotic, dissipative, 3-variable ODE systems.
 The algorithms encompass three approaches to
 exploiting the temporal structure of the input. 
  
 Our results indicate that the best algorithm depends on the dynamical
 system, the size of the data set, and on the number of time steps included in the input.
  Overall, the results show that in certain conditions, the LLEs can be predicted well:
 this depends on the system dynamics, the LLE being predicted,
 and importantly on the local heterogeneity of the LLE in the proximity of the
 given state.
 In particular, the average accuracy was lowest for the neutral
 LLE. Further work is required to see if this result also holds in
 ocean-atmosphere systems (and multiscale systems more generally), where the
 neutral and near-neutral exponents are key to determining local predictability
 \citep{DeCruz2018, Quinn2020}.
Our results suggest that the feasibility of using supervised ML to drive adaptive actions in an operational setting will depend on the specific
 use case: the forecasting model, the desired target information, and the
 intended adaptive actions. 
 
 Additionally, we investigated the impact of the size of the data set used to
train the ML algorithms. We found that with data sets of $10^6$ examples,
compared to $10^5$ examples, variance of \rs score reduced but there
were only marginal improvements in the mean \rs score. With the $10^6$ data
sets, the LSTM performed best in both systems. However, with the $10^5$ data
set, the LSTM was limited: the MLP performed best in the \ros whereas the CNN
performed best in the \lst.
 The RT achieves accuracy that is close to the best-performing algorithm in both
 systems, whilst being
computationally much cheaper than the NNs.
 We tested two input types: one with $1$ time step, and one with $6$ time steps (of the system state). We found the best input type depends on data set size and dynamical system: in the \ros, $6$ input time steps is better for the smallest data set ($5\times 10^4$), whilst in the \lst, much more data ($10^6$)  is required for the LSTM and MLP to achieve comparable performance with the $6$-time step input.
 We further show that large prediction errors occur when the current state is in a region of local heterogeneity on the system attractor.
 Outside of the locally heterogeneous regions, the best-performing algorithms make consistently accurate predictions. 
The differences in local heterogeneity between the two systems explain the lower \rs scores achieved for LLEs 1 and 2 in the \ros, compared to the \lst. 
We explain that local heterogeneity is an insurmountable problem for deterministic ML predictions.
This challenge could be mitigated if the ML prediction also included a
reliable uncertainty quantification.
We suspect that an uncertainty quantification could be made either by using
Bayesian NNs \citep{Wang2016} or by including a measurement of the nearby local
heterogeneity in the target of each example.

The low-dimensional setting permitted extensive experimentation in this work, 
providing lessons that will be useful should this ``non-intrusive'' approach be taken in weather
and climate prediction.
The next steps will be to apply the approach of this work to spatially-extended
models with more dimensions.
There are several foreseeable challenges on the path from
the very low-order models of this work to the envisioned setting of operational
weather prediction models.
The first challenge is to  generate suitable data sets, since the calculation of LLEs does not scale well and requires a tangent linear model (see \s\ref{subsec computation
method}).
However, the requirement of a tangent linear model can be avoided by using bred vectors \parencite[\eg][]{Toth1997,Uboldi2015}.
Additionally, the attractor of any numerical weather prediction (NWP) model is complex and high-dimensional: very large data sets
will be required if sufficient attractor coverage is to be obtained.
For a NWP model with $\O(10^8)$ variables, one would need $\O(10^{12})$ input-target
pairs to obtain a ratio between the degrees of freedom of the NWP model and the
size of the ML data set that is similar to the ratio used in this study.
If one were to use ERA5 reanalysis data \citep{era5} as input, a data set of
$\O(10^{12})$ single time step inputs would amount to approximately
$\O(10^{9})$~TB of data, which is unfeasibly large. Furthermore,
due to the long time scales involved in teleconnection events, the number of
such events can be small even in long time-series. Given the number of input
features, this can lead to a ``small data problem'', see \eg \textcite{Vecchi2022}
and references therein.

Therefore, it may be necessary to generate training data using a reduced-dimension version of the
operational model. For example, in
\textcite{Quinn2021,Quinn2022}, LLEs are computed by reducing the data dimension
(via empirical orthogonal functions) and constructing a multi-state vector
auto-regressive model.

Once initial training data has been generated, the cost of making
predictions with the ML model can be reduced by further reducing the dimension
of the training data, \ie by performing feature-extraction \parencite{Guyon2006}.
For example, ML techniques such as autoencoders can be used for dimension
reduction, \eg see \textcite{Mack2020}. Also, it may be possible to curate
training data sets strategically to reduce their size. 
Finally, if the intended use-case requires only part of the LLE spectrum, then
cost savings can be made 
\textit{a)} when generating training data, which scales as $\O(n^2)$ rather than
$\O(n^3)$ (see \s\ref{sec computational cost}), and \textit{b)} when making
predictions (see \s\ref{sec computation time}).

The computational benefit of the ML approach investigated here is two-fold. 
The ML approach estimates LLEs directly from the current system state, thus
avoiding the cost of the long spin up that is required by the conventional method for
calculating LLEs.
Second, the ML approach has the potential to be cheaper per iteration of LLEs. 
We found that the lightest algorithm we tested, the RT, was computationally
cheaper (by a factor of 10) than the conventional method for calculating LLEs
(see \tab\ref{table computational cost}).

Although the NNs were comparatively costly in this setting, we expect that in a
higher-dimensional,
operational setting, NNs may be competitive.
It is unknown how the required NN size will increase with the
system dimension: this will require experimentation.
The time-cost of making predictions with NNs may be reduced (relative to the size
of the NNs) by using purpose-built ML hardware.
On the other hand, the cost of numerically calculating LLEs (by propagating perturbations and orthogonalising) will scale as
$\O({n^3})$ if computing the full spectrum, or $\O({n^2})$ if the number of LLEs
computed is much smaller than the dimension of the NWP model $n$.

\section*{Acknowledgements}
DA is funded by a studentship from the Engineering and Physical Sciences Research Council (EP/N509723/1). JA and AC acknowledge the support of the UK National Centre for Earth Observation (grant no. NCEO02004). AC is also supported by the project SASIP funded by Schmidt Futures – a philanthropic initiative that seeks to improve societal outcomes through the development of emerging science and technologies.

\printbibliography[title=References]

\end{document}